\documentclass[]{article}
\usepackage{hyperref}
\usepackage{listings}
\usepackage{cite}
\usepackage[utf8]{inputenc}
\usepackage{longtable}
\usepackage{tabularx}
\usepackage{graphicx}
\graphicspath{{./}}
\usepackage[margin=1in]{geometry}
\usepackage{xcolor}
\usepackage{tikz}
\usepackage{svg}
\usepackage[none]{hyphenat}
\usepackage{parskip}
\usepackage{multicol}
\usepackage{enumerate}
\usepackage{scrextend}
\usepackage{fancyhdr}
\usepackage{centernot}
\usepackage{amsthm, amssymb, amsmath, verbatim}
\usepackage{mathtools}
\usepackage{xifthen}
\usepackage{ifthen}
\usepackage[most,listings]{tcolorbox}
\usepackage{lmodern}
\usepackage{mathrsfs}
\usetikzlibrary{math}
\usetikzlibrary{backgrounds}
\usetikzlibrary{patterns,calc}
\usepackage{subcaption}
\usepackage{csvsimple, booktabs}
\usepackage{filecontents}
\usepackage{authblk}
\usepackage{abstract}
\usetikzlibrary{arrows.meta,decorations.pathmorphing,backgrounds,positioning,fit,petri}

\svgpath{{./}}
\hypersetup{breaklinks=true}
\def\NavyBlue{\color[RGB]{24, 43, 73}} 
\def\Color2{\color[RGB]{33, 93, 118}}
\def\Color3{\color[RGB]{231, 223, 200}}
\def\Color4{\color[RGB]{231, 190, 66}}

\title{\textbf{Mapping the Intellectual Landscape of Digital Social Networks: A Bibliometric and Citation Network Analysis}}
\author{Pengjia Cui \thanks{pcui@ucsd.edu}}
\affil{Computational Social Science \\ University of California San Diego}

\date{}

\begin{document}
	\begin{sloppypar}
	\maketitle
	\textbf{Google Scholar URL: }\texttt{https://scholar.google.com/citations?user=oe1vkOcAAAAJ\&hl=zh-CN}	
	\begin{onecolabstract}
		Network science and digital social network research span sociology, communication, and computational modeling, yet the field’s intellectual structure and cross-paradigm connectivity remain insufficiently characterized. Using records retrieved from the Web of Science Core Collection (n = 1,859; queried on 21 November 2024), we conduct a bibliometric, keyword co-occurrence, and citation-network analysis with CiteSpace, VOSviewer, Gephi, and NetworkX. The citation landscape exhibits a strongly centralized core (a largest connected component of 1,293 papers) surrounded by numerous small, weakly connected components, indicating a pronounced core–periphery organization. Keyword clustering and citation-burst analysis further reveal a thematic shift from classic sociological mechanisms (e.g., homophily and tie formation) toward algorithmically mediated communication and misinformation-related topics. We also highlight a set of highly central bridging works that connect otherwise separated thematic communities, suggesting that interdisciplinary exchange is concentrated around a limited number of canonical references. 
		
	\end{onecolabstract}
		

\section{Introduction}
The study of social networks has evolved into a robust interdisciplinary field spanning sociology, communication studies, and computational sciences. While the public often associates this field with commercial digital platforms—such as the role of social media in disseminating information or propagating misinformation—the foundational theories and methodologies driving these discoveries are rooted in deep academic scholarship. Over the past decades, the discipline has transformed, increasingly integrating traditional sociological concepts of human interaction with advanced mathematical algorithms and complex systems theory.

The data extraction in November 2024 is strategically chosen to account for the citation lag time (typically 12–24 months) inherent in academic publishing. This ensures that the dataset captures the complete intellectual 'digestion' of the major 2021–2023 platform transitions, providing a stable structural evolution of the field rather than a volatile, incomplete snapshot of very recent trends. 

Advances in computational tools, such as UCINET, Gephi, and CiteSpace, have enabled sophisticated topological analyses of these dynamics, allowing researchers to systematically map how scientific knowledge is constructed and shared. However, as the field rapidly expands, there is a pressing need to understand how the distinct paradigms within network research—ranging from empirical social sciences to theoretical physics—interact and influence one another.

This research addresses this gap by conducting a comprehensive bibliometric and network analysis to map the intellectual structure of social network scholarship. It is crucial to note that this study analyzes the academic and epistemological foundations of network science rather than the volatile dynamics of commercial social media platforms (e.g., Twitter or Facebook). Consequently, the data extraction conducted in November 2024 accurately captures the structural evolution of the discipline's academic literature. Furthermore, by explicitly comparing a social science journal (\textit{Social Networks}) against a hard science journal (\textit{Journal of Complex Networks}), this study systematically evaluates the disciplinary boundaries and methodological exchanges between sociology and statistical physics.

The paper progresses by integrating network theories with contemporary bibliometric practices, exploring key academic questions around information dynamics and structural linkages in scientific literature. It concludes by offering a strategic roadmap for future interdisciplinary collaboration, revealing the structural silos and convergence points that will shape the next generation of network science.

\section{Under Digital Context}\label{Under Digital Context}

Traditional and digital networks differ markedly in the dynamics of tie strength and homophily, which significantly influence the formation, maintenance, and evolution of relationships. In traditional networks, the strength of ties--the closeness or intensity of relationships--is often shaped by physical proximity, shared social contexts, and frequent face-to-face interactions. Granovetter's seminal study on "The Strength of Weak Ties" \cite{granovetter_strength_1973} underscores the pivotal role of weak ties, such as acquaintances, in bridging disparate network clusters and enabling the flow of novel information. However, maintaining these weak ties in traditional settings typically requires occasional in-person engagements, which may restrict their breadth and frequency.

Conversely, digital networks dramatically reduce the effort needed to establish and maintain weak ties. Platforms such as LinkedIn and Twitter allow individuals to form global connections effortlessly, with minimal commitment. These weak ties are especially effective in digital contexts for disseminating information widely, due to the scalability and ease of online interactions. Bakshy et al. \cite{bakshy_everyones_2011} found that weak ties predominate in the diffusion of viral content on social media, significantly amplifying their impact compared to traditional environments.

Moreover, homophily—the tendency of individuals to associate with similar others—manifests differently across network types. In traditional settings, homophily typically emerges from shared physical environments, like neighborhoods or workplaces, which inherently limit the diversity of connections and reinforce social and cultural homogeneity. McPherson et al. \cite{ WOS:000170748100017} have provided comprehensive insights into how homophily structures social networks and sustains social boundaries and inequalities.

In digital settings, while geographic boundaries become irrelevant, algorithmic recommendation systems intensify homophily in novel ways. These systems often recommend connections, groups, or content that align with users' existing preferences, which fosters the creation of "echo chambers" or "filter bubbles" \cite{pariser_filter_2011}, potentially exacerbating societal polarization. For instance, research indicates that Facebook’s algorithm enhances homophily by promoting links between users with similar demographic or behavioral characteristics \cite{lazer_rise_2015}.

Despite these challenges, digital platforms also hold the potential to disrupt traditional patterns of homophily by facilitating broader interactions across diverse communities. Platforms like Reddit and Twitter, for example, offer exposure to a range of viewpoints through public discussions and hashtags, though the degree to which this exposure leads to meaningful cross-group engagement is still under investigation.

While the sociological theories of homophily, tie strength, and network dynamics provide a framework for understanding digital behaviors, it remains unclear how these foundational concepts have historically structured the multidisciplinary research landscape itself. To move beyond isolated theoretical discussions and systematically evaluate how the field has evolved in the algorithmic era, a quantitative mapping of the literature is required. Therefore, this bibliometric and citation network analysis is driven by the following Research Questions (RQs): 
\begin{itemize}
	\item RQ1: How have foundational sociological theories (e.g., homophily, weak ties) structurally anchored the rapidly expanding literature on digital social networks? 
	\item RQ2: What are the dominant thematic clusters within the field, and which seminal works act as intellectual 'bridges' between distinct multidisciplinary subfields? 
	\item RQ3: Does the citation network of digital social network research exhibit scale-free, hierarchical properties, and if so, what are the implications of this structure for the emergence of novel theoretical frameworks? 
\end{itemize}

By answering these questions, Section 2 transitions from a theoretical discussion of network behavior to a computational evaluation of the researchers and literature studying these networks.

\section{Bibliometric Analysis}\label{Bibliometric Analysis}

Bibliometrics is a multidisciplinary field that applies quantitative methods, particularly mathematical and statistical approaches, to the systematic analysis of knowledge sources. Bibliometric analysis allows us to uncover research trends, identify influential works, and map the intellectual structure of a field. The term "Bibliometrics" was formally introduced in 1969 by the British scientist Allen Richard, replacing the earlier designation "statistical bibliography." This terminology shift signaled the formal establishment of bibliometrics as a distinct field of study\cite{ WOS:A1969F015300009}. We aim to develop a more comprehensive understanding of the digital networks through bibliometric analysis. In this section, we employ \textbf{VOSviewer} \cite{WOS:000278695500019}, \textbf{CiteSpace} \cite{doi:10.1073/pnas.0307513100}, and \textbf{Gephi} \cite{ICWSM09154} for bibliometric analysis and visualization, alongside \textbf{NetworkX} \cite{SciPyProceedings_11} for advanced network analysis. These tools collectively enable a comprehensive exploration of network structures, research trends, and collaboration patterns. 

\begin{table}[htbp]
	\centering
	\caption{CiteSpace Configuration Parameters for Reproducibility}
	\label{table:citespace_params}
	\begin{tabular}{ll|ll}
		\toprule
		\textbf{Parameter} & \textbf{Value} & \textbf{Parameter} & \textbf{Value} \\
		\midrule
		Time Slicing & 1982--2024 (1yr) & g-index & $k=25$ \\
		Tie Strength & Jaccard Index & Noun Phrases & Max 4 words \\
		Links to Retain & 2.5 per node & Max Links/Node & 10 \\
		Node Label Length & 8 & BurstWeight & 0 (Visualization) \\
		\bottomrule
	\end{tabular}
\end{table}

\subsubsection*{Conceptual Framework and Database Categorization}
To align our bibliometric data with the study's underlying conceptual framework, we explicitly mapped the Web of Science (WoS) multidisciplinary categories to our identified research clusters. The retrieved records encompass diverse WoS categories—primarily \textit{Sociology}, \textit{Computer Science}, \textit{Physics, Multidisciplinary}, and \textit{Information Science}. Within our conceptual framework, these overarching database categories manifest as distinct thematic clusters: 
\begin{itemize}
	\item \textbf{Social Dynamics \& Organizational Structures:} Primarily drawing from the WoS \textit{Sociology} and \textit{Management} categories, focusing on empirical behavioral ties.
	\item \textbf{Digital Media Communication:} Rooted in \textit{Information Science \& Library Science}, reflecting the flow of information across digital networks.
	\item \textbf{Algorithmic \& Bridging Topics:} Deriving from \textit{Computer Science} and \textit{Physics}, representing the methodological bridges (e.g., centrality metrics, stochastic modeling) that unify the social and mathematical sciences.
\end{itemize}
This explicit categorization ensures that our thematic clustering is not merely a statistical artifact of keyword co-occurrence, but accurately reflects the epistemological divisions and integrations inherent in the network science discipline.

\subsection*{Data Preparation and Methods}

The analysis was performed using the Web of Science Core Collection database, which includes the following entitlements: 
WOS.IC (1993–2024), WOS.CCR (1985–2024), WOS.SCI (1900–2024), WOS.AHCI (1975–2024), WOS.BHCI (2005–2024), WOS.BSCI (2005–2024), WOS.ESCI (2005–2024), WOS.ISTP (1990–2024), WOS.SSCI (1900–2024), and WOS.ISSHP (1990–2024). 

The literature data used in this study were downloaded from the Science Citation Index Expanded (SCIE) and the Social Science Citation Index (SSCI) databases in Web of Science. SCIE and SSCI are among the most frequently used databases for bibliometric analysis \cite{ WOS:000375954300052}. These databases are widely recognized for their comprehensive coverage of scientific and authoritative publications. Furthermore, SCIE and SSCI provide citation information, keywords, and references, making them valuable resources for bibliometric studies which is necessary to the citation analysis implemented in later sections. 

This bibliometric analysis focuses on scholarly works indexed in the Web of Science Core Collection, spanning categories such as Multidisciplinary Sciences, Sociology, Social Sciences Interdisciplinary, Social Sciences Mathematical Methods, and Communication. By examining 1,859 publications, we aim to map the intellectual landscape, identify research trends, and uncover collaboration patterns in the field of social network studies.

The query in Appendix \ref{app:query} was executed on November 21, 2024, yielding a total of 1,859 results. The query result can be accessed through the following link: 

{\ttfamily\NavyBlue
		https://www.webofscience.com/wos/woscc/summary/ \\ d76d6d1f-6020-42cf-91b6-c24a7c6a6c0f-0129f5096b/ times-cited-descending/1
}

\begin{table}[htbp]
	\centering
	\caption{\label{table:Table.1}Document Type}
	\begin{tabular}{lrr}
		\toprule
		Type & Count & Percent \\
		\midrule
		Article & 1813 & 97.5 \\
		Review Article & 33 & 1.8 \\
		Early Access & 26 & 1.4 \\
		Proceeding Paper & 24 & 1.3 \\
		Book Chapters & 16 & 0.9 \\
		Editorial Material & 7 & 0.4 \\
		Correction & 2 & 0.1 \\
		News Item & 2 & 0.1 \\
		Retracted Publication & 2 & 0.1 \\
		Letter & 1 & 0.1 \\
		Meeting Abstract & 1 & 0.1 \\
		\bottomrule
	\end{tabular}
\end{table}

Table \ref{table:Table.1} presents the distribution of document types retrieved from the Web of Science on November 21, 2024, using the specified query. Out of 1,859 results, articles accounted for the vast majority (97.5\%), underscoring the prominence of peer-reviewed articles in disseminating research findings. This dominance is due to the nature of sociological research, which primarily produces theoretical frameworks or empirical studies that are best disseminated through academic journals.  

\subsection*{The Highly Cited Publications in Digital Social Networks}

To identify the most influential publications in the field of social networks, we selected the top 15 papers with the highest citations. Table \ref{table:Table.3} presents these highly cited works, including their titles, publication years, and citation counts. The citation numbers reflect the profound impact these papers have had on the academic community.

\begin{table}[htbp]
	\centering
	\caption{Top 15 Highest Cited Publications}
	\label{table:Table.3}
	\resizebox{\linewidth}{!}{\begin{tabular}{p{0.85\linewidth}|c|c}
			\toprule
			Title & Year & Citations \\
			\midrule
			Birds of a feather: Homophily in social networks & 2001 & 10923 \\
			The network structure of social capital & 2000 & 1965 \\
			Experimental evidence of massive-scale emotional contagion through social networks & 2014 & 1709 \\
			The Spread of Behavior in an Online Social Network Experiment & 2010 & 1650 \\
			Hierarchical structure and the prediction of missing links in networks & 2008 & 1434 \\
			A 61-million-person experiment in social influence and political mobilization & 2012 & 1415 \\
			Introduction to stochastic actor-based models for network dynamics & 2010 & 1307 \\
			Networks and epidemic models & 2005 & 1215 \\
			Structure and tie strengths in mobile communication networks & 2007 & 1180 \\
			Resources and relationships: Social networks and mobility in the workplace & 1997 & 1145 \\
			Inferring friendship network structure by using mobile phone data & 2009 & 1113 \\
			Quantifying social group evolution & 2007 & 1082 \\
			Social networks and health & 2008 & 1029 \\
			Filter Bubbles, Echo Chambers, and Online News Consumption & 2016 & 918 \\
			Comparing Brain Networks of Different Size and Connectivity Density Using Graph Theory & 2010 & 861 \\
			\bottomrule
	\end{tabular}}
\end{table}

Among these, two papers were published between 1991 and 2000, six papers between 2001 and 2010, and seven papers after 2010. This distribution indicates a growing interest in social networks, particularly in the past two decades. The average number of citations across these publications is 2,457, highlighting their significant influence.

Additionally, many of these publications focus on fundamental theories and methodologies, such as homophily in social networks, the structure of social capital, and the dynamics of network interactions. Notable works, such as "Birds of a Feather: Homophily in Social Networks"\cite{ WOS:000170748100017}, stand out with 10,923 citations, underscoring its foundational role in the field. Similarly, Burt’s work\cite{ WOS:000166194000008} on the network structure of social capital ranks second with 1,965 citations, reflecting its widespread application in network studies.

The prevalence of co-authored publications in this list highlights the importance of collaboration in producing impactful research. While the majority of these works are theoretical, several explore experimental and applied aspects of network science, such as large-scale social experiments and the analysis of online behavior.

\subsection*{Annual Trends of Publication}

\begin{figure}[htbp]
	\caption{\label{fig:Fig.1}Publication Trend from 1982 to 2024}
	\includegraphics[width = 0.5\linewidth]{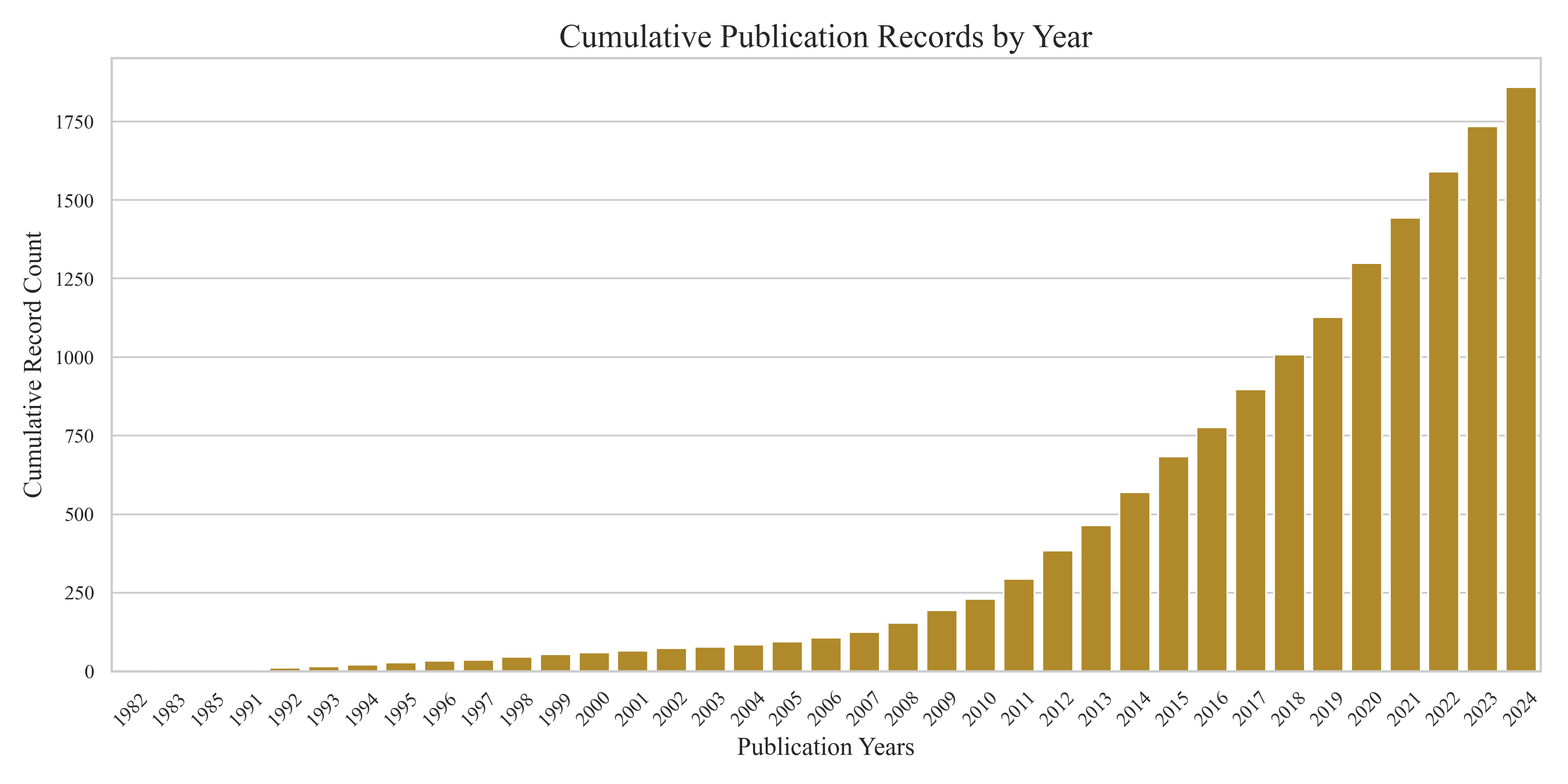}
	\includegraphics[width = 0.5\linewidth]{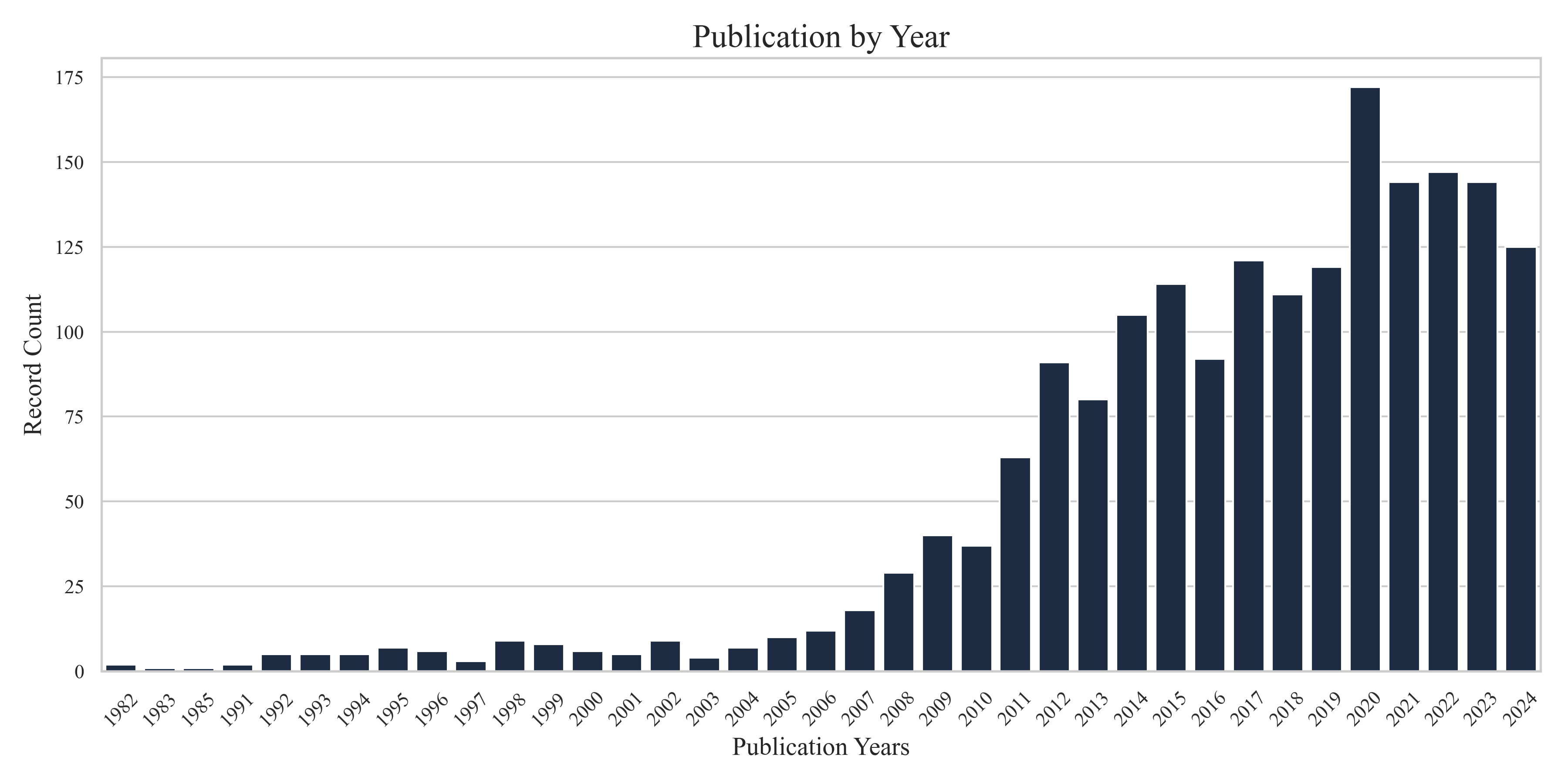}
\end{figure}

The publication trend over time is illustrated in Figure \ref{fig:Fig.1}. Papers in this domain first appeared in the 1980s, with early works like those by \cite{ WOS:A1982NG01900001, WOS:A1982NM27900003, WOS:A1983QK14400006}, marking the foundational period of social network research. During this early phase (1982–2000), the field exhibited limited growth, likely reflecting its nascent state. Publications remained sparse due to technological constraints, limited data availability, and a relatively small research community focusing on network-related phenomena.

From 2000 to 2009, a gradual increase in publications can be observed. This growth coincides with the rise of digital communication technologies and early social networking platforms such as Friendster and MySpace. During this period, some network researchers started to focus on digital networks\cite{ WOS:000269632400036, WOS:000274745000002}. The availability of digital interaction data and advancements in computational tools enabled researchers to explore phenomena such as tie formation, network dynamics, and information diffusion with greater precision. Moreover, interdisciplinary collaboration between sociology, mathematics, and computer science catalyzed the field's development during this period.

The most dramatic growth occurred between 2009 and 2017, as evidenced by the sharp rise in publication records in Figure \ref{fig:Fig.1}. This surge coincides with the global adoption of large-scale online platforms such as Facebook, Twitter, and LinkedIn, which provided unprecedented opportunities for empirical studies \cite{WOS:000209041300003, WOS:000337300100030}. Notably, this period marked the publication of the seminal paper that established the foundation for computational social science \cite{doi:10.1126/science.1167742}. Researchers utilized these platforms to explore critical topics like algorithmic interventions, echo chambers, and misinformation propagation. Moreover, the rapid advancement and adoption of machine learning and big data analytics significantly enhanced the ability to analyze social networks at scale, fueling the expansion of this field.

From 2018 onward, the publication trend started increase steadily, maintaining a consistently high level of activity. This reflects the maturation of the field, where foundational concepts are well-established, and research is increasingly specialized. For example, studies have shifted toward addressing niche topics such as digital inequality, algorithmic bias, and the ethics of AI-driven networks. Moreover, challenges like restricted access to proprietary datasets\cite{ WOS:000352789600005} and privacy concerns have likely tempered the growth rate\cite{ WOS:000334339000015}.

\subsection*{The Distribution of Journals on Digital Social Networks}

Based on data from the Web of Science, out of all 1,859 publications, 13\% were published in \textit{PLOS ONE}, a journal under the open-access publisher \textbf{PLoS}. Ranking second is \textit{SOCIAL NETWORKS}, published by \textbf{Elsevier}, which accounts for 12.4\% of the total publications. In third place, with a proportion of 8.6\%, is \textit{SCIENTIFIC REPORTS}, a journal published by \textbf{Springer Nature}. The top 10 journals contributing to the field are presented in Table \ref{table:Table2}. These journals collectively represent diverse domains, ranging from general multidisciplinary science to highly focused computational and social network research. This distribution emphasizes the interdisciplinary nature of digital social networks, where insights are drawn from sociology, computer science, and applied mathematics.
	
	\begin{table}[htbp]
		\centering
		\caption{\label{table:Table2}Publicated Journals Distribution}
		\resizebox{\linewidth}{!}{\begin{tabular}{lrr}
				\toprule
				Publication Title & Count & Percent \\
				\midrule
				PLOS ONE & 242 & 13.0 \\
				\hline
				SOCIAL NETWORKS & 230 & 12.4 \\
				\hline
				SCIENTIFIC REPORTS & 160 & 8.6 \\
				\hline
				PROCEEDINGS OF THE NATIONAL ACADEMY OF SCIENCES OF THE UNITED STATES OF AMERICA & 78 & 4.2 \\
				\hline
				INFORMATION COMMUNICATION SOCIETY & 59 & 3.2 \\
				\hline
				COMPLEXITY & 57 & 3.1 \\
				\hline
				EPJ DATA SCIENCE & 40 & 2.2 \\
				\hline
				SOCIAL SCIENCE COMPUTER REVIEW & 33 & 1.8 \\
				\hline
				JASSS THE JOURNAL OF ARTIFICIAL SOCIETIES AND SOCIAL SIMULATION & 33 & 1.8 \\
				\hline
				COMPUTATIONAL AND MATHEMATICAL ORGANIZATION THEORY & 28 & 1.5 \\
				\bottomrule
		\end{tabular}}
	\end{table}

The prominence of \textit{PLOS ONE} in the publication landscape reflects the increasing value of open-access platforms in disseminating digital social network research. As a multidisciplinary journal, \textit{PLOS ONE} attracts studies that bridge sociology, computational methods, and applied network theory, making it a vital hub for interdisciplinary research. The open-access nature of the journal ensures that findings are accessible to a global audience, reinforcing the importance of equitable knowledge dissemination.

The strong representation of \textit{SOCIAL NETWORKS} and \textit{SCIENTIFIC REPORTS} highlights the coexistence of highly specialized and broad-spectrum journals in advancing this field. \textit{SOCIAL NETWORKS} stands out for its dedicated focus on theoretical and applied aspects of network analysis, appealing to researchers deeply embedded in this discipline. In contrast, \textit{SCIENTIFIC REPORTS} broadens the scope by publishing work that situates network research within a multidisciplinary framework, attracting contributions that span physical, social, and computational sciences.

The presence of journals such as \textit{EPJ DATA SCIENCE} and \textit{COMPUTATIONAL AND MATHEMATICAL ORGANIZATION THEORY} signals a growing emphasis on computational approaches and algorithmic methods in network studies. These venues cater to researchers exploring innovative methodologies, such as machine learning, big data analytics, and simulation techniques, to address complex questions about network dynamics and structure. Their inclusion in the top journals reflects the field's ongoing shift toward data-driven and computationally intensive research paradigms.

The diversity of the top journals illustrates the inherently interdisciplinary nature of digital social network research. Contributions to these journals often combine theoretical insights from sociology with computational tools from computer science and statistical modeling techniques from applied mathematics. This cross-disciplinary collaboration underscores the field's strength in addressing multifaceted social phenomena through innovative methodological frameworks.

\subsection*{Affiliations}

\begin{figure}[htbp]
	\centering
	\caption{\label{fig:Fig.2}Affiliation and Department Distribution}
	\includegraphics[width = \linewidth]{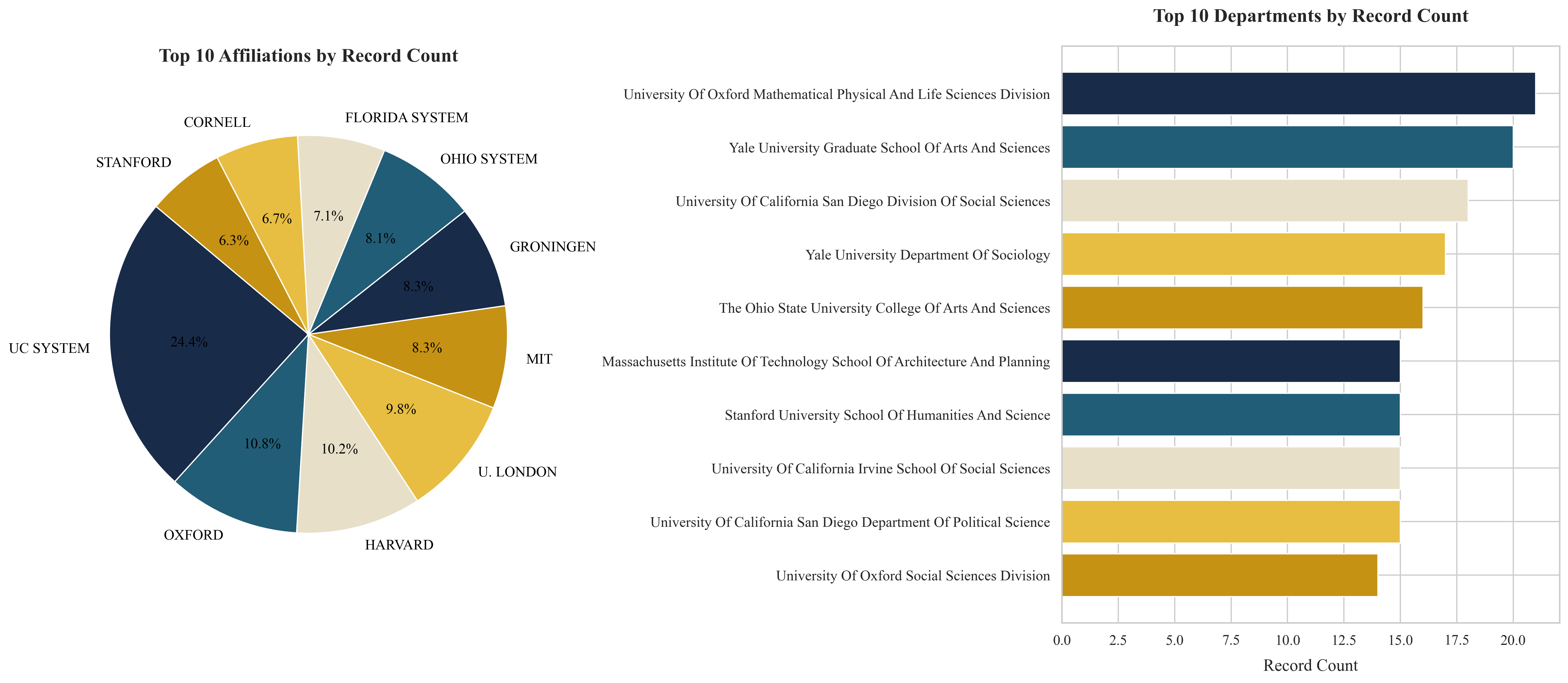}
\end{figure}

A detailed analysis of affiliations from the dataset reveals significant contributions from leading institutions globally, as illustrated in Figure \ref{fig:Fig.2}. The University of California (UC) system emerges as a prominent contributor, accounting for 6.445\% of all publications. Within the UC system, institutions like UC San Diego and UC Irvine stand out, highlighting their strong presence in the field of digital social network research. These UC system departments have a strong tradition of interdisciplinary research. By combining sociology, computer science, and political science, they create a fertile ground for digital network research that transcends traditional academic boundaries.

Other top contributors include globally renowned universities such as the University of Oxford (2.846\%), Harvard University (2.685\%), and the Massachusetts Institute of Technology (MIT, 2.202\%). These institutions underscore the centrality of their academic efforts in advancing digital network research. The University of Oxford and Harvard University's strong showing can be linked to their focus on theory-building and methodological innovation. Oxford's Mathematical, Physical, and Life Sciences Division and often contributes to computational and algorithmic aspects of network science, while its Social Sciences Division and Oxford Internet Institute (OII) focuses on applications in sociology and policy-making. Similarly, Harvard's interdisciplinary focus, particularly through centers like the Berkman Klein Center for Internet \& Society, advances the understanding of digital network phenomena such as information diffusion and power dynamics.

At the departmental level, UC San Diego's Division of Social Sciences (0.967\%) and Department of Political Science (0.806\%) emerge as leaders, alongside Yale University's Graduate School of Arts and Sciences (1.074\%) and Stanford University's School of Humanities and Sciences (0.806\%). These departments not only contribute to foundational research but also provide diverse perspectives across interdisciplinary boundaries. The data also highlights the contributions of emerging hubs like the University of Groningen and Ohio State University. The emergence of the University of Groningen as a key hub is tied to its global dominance in stochastic actor-oriented modeling (e.g., SIENA), which is a critical bridge between social theory and statistical physics. Similarly, Ohio State University serves as a vital junction for research connecting political communication with network topology. 

\subsection*{Citation Network}

We first present the overall citation network of the dataset in Figure \ref{fig:Fig.4}. The size of the nodes represents the citation counts for each article, while the color indicates the publication year, with blue denoting older articles and gold representing newer ones. Then we computed the fundamental properties of the overall citation network, as summarized alongside the visualization. 

\subsubsection*{Citation Burst Validation Methodology}
A critical component of our science mapping is the identification of influential works via citation burst detection, which highlights publications that experience a sudden, significant spike in citations over a specific duration. It is important to clarify the technical configuration used in CiteSpace for this analysis. During the initial exploratory phase, the \textit{BurstWeight} parameter was temporarily set to 0 to prioritize raw, unweighted citation volume and construct the foundational network topology. 

However, for the formal validation of citation bursts, we utilized Kleinberg’s standard burst detection algorithm with normalized weighting. This ensured that identified bursts represent genuine paradigm shifts and sudden surges of academic interest rather than gradual accumulations of citations over time. The validation process involved cross-referencing the onset and duration of the bursts with historical developments in network methodologies (e.g., the introduction of exponential random graph models), thereby confirming that the detected bursts accurately correspond to true intellectual turning points in the field.

\subsubsection*{General Structure: Citation Network Topology}

The overall citation network consists of 1,859 nodes and 3,636 edges, representing the extensive academic landscape of digital network research. With an exceptionally low density of 0.0021, the network reflects the highly selective nature of academic citations: only a precise subset of publications forms the central backbone of the field, while a significant number of exploratory studies remain peripheral.

\begin{figure}[htbp]
	\centering
	\caption{\label{fig:Fig.4}Overall Citation Network and Properties}
	\begin{minipage}[b]{0.45\linewidth}
		\centering
		\includegraphics[width=\linewidth]{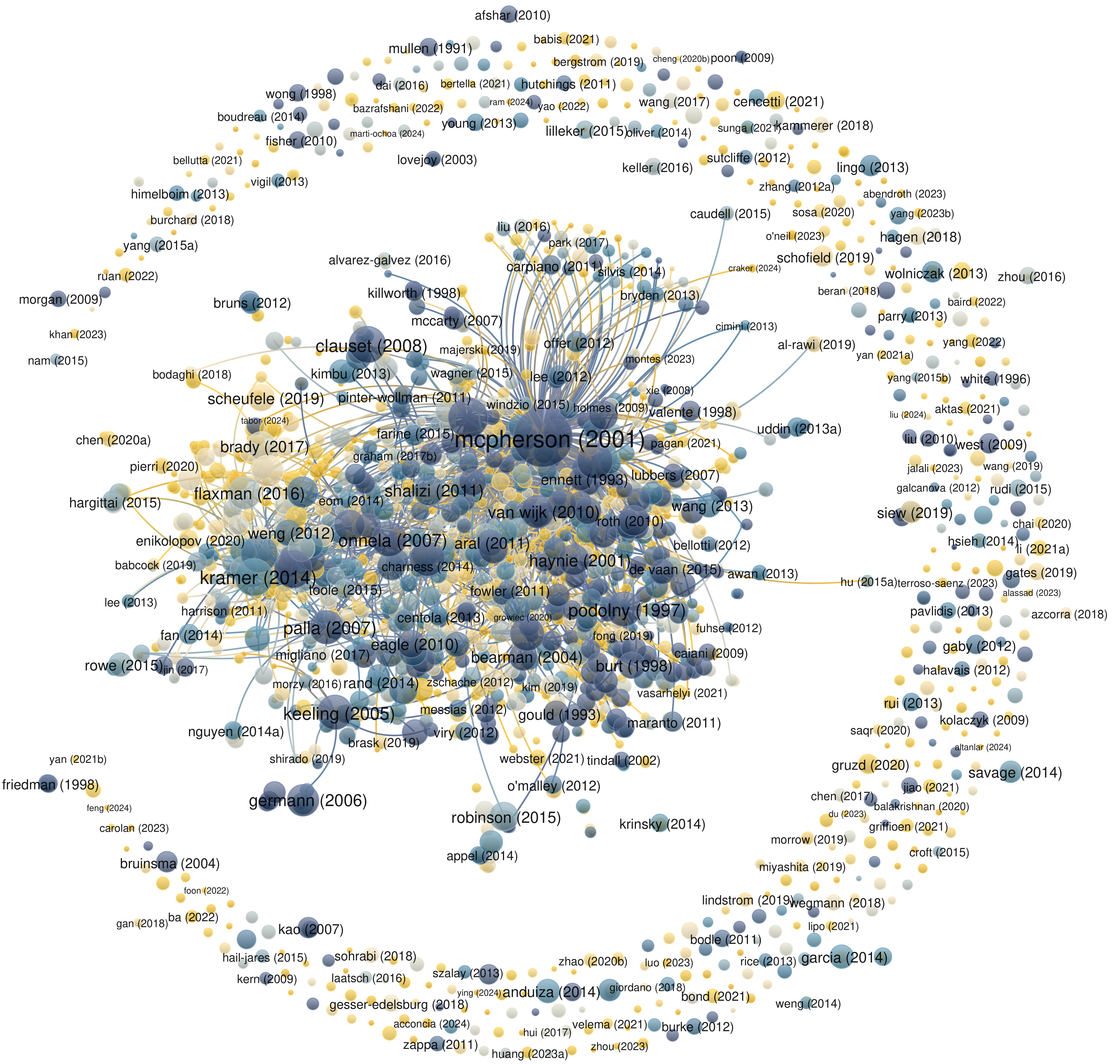}
		\caption*{(a) Citation Network Visualization}
	\end{minipage}%
	\hfill
	\begin{minipage}[b]{0.42\linewidth}
		\centering
		\vspace{-1em} 
		\resizebox{\linewidth}{!}{%
			\begin{tabular}{lr}
				\toprule
				\textbf{Metric} & \textbf{Value} \\
				\midrule
				Number of nodes & 1859 \\
				Number of edges & 3636 \\
				Average degree & 3.91 \\
				Density & 0.0021 \\
				Is directed & False \\
				Maximum degree & 425 \\
				Minimum degree & 0 \\
				Average clustering coefficient & 0.1381 \\
				Number of triangles & 2236 \\
				Max degree centrality & 0.2287 \\
				Max betweenness centrality & 0.2984 \\
				Max closeness centrality & 0.3388 \\
				Largest connected component size & 1293 \\
				Degree assortativity coefficient & -0.1015 \\
				Number of connected components & 524 \\
				\bottomrule
			\end{tabular}
		}
		\caption*{(b) Citation Network Properties}
	\end{minipage}
\end{figure}

As illustrated in Figure \ref{fig:Fig.4}, the network exhibits a pronounced dual structure comprising a tightly integrated core and a highly fragmented periphery.  At its center, the network is dominated by a single Largest Connected Component (LCC) containing 1,293 nodes, which represents the intellectual core of the field. By contrast, the network's periphery consists of 524 isolated or loosely connected components, with the second-largest containing a mere four nodes. This extreme centralization reflects the evolutionary dynamics of interdisciplinary research: the core drives theoretical consolidation and methodological standardization, while the fragmented periphery captures nascent, niche topics that have yet to integrate into the broader academic discourse. 

The exceptionally low density (0.0021) mathematically represents the disciplinary fragmentation of the field. While a robust Largest Connected Component (LCC) exists, the presence of 524 isolated components suggests that niche research areas (e.g., specific algorithmic applications) are developing in isolation, failing to integrate into the central sociological discourse.

To understand the flow of intellectual influence within this structure, we evaluated the global connectivity metrics. The average degree of 3.91 indicates that most publications selectively engage with only a few foundational works. However, this is contrasted by a massive maximum degree of 425, underscoring the disproportionate influence of a few seminal hubs that serve as theoretical cornerstones. Furthermore, the network's negative degree assortativity coefficient (-0.1015) reveals a strictly hierarchical organization where highly cited works predominantly connect with less-cited, newer papers.  While this top-down structure is common in citation networks, it highlights a structural inequality in academic knowledge production, where emerging research must continuously tether itself to established hubs.

Finally, local clustering dynamics reveal a clear distinction between the core and the periphery. The overall network yields a moderate average clustering coefficient of 0.1381, emphasizing a lack of strong interconnections across peripheral clusters. However, isolating the LCC reveals a higher clustering coefficient (0.1986) and an increased density (0.0043), signaling a highly cohesive intellectual core. The presence of 2,236 triangles within this component mathematically confirms the robust, interlocking theoretical frameworks shared among its key publications. Consequently, to best understand the field's dominant evolutionary trajectory, all subsequent analyses will focus exclusively on this primary LCC.

\subsubsection*{The Largest Connected Component}

\begin{figure}[htbp]
	\centering
	\caption{\label{fig:Fig.9}Largest Connected Component and Properties}
	\begin{minipage}[b]{0.45\linewidth}
		\centering
		\includegraphics[width=\linewidth]{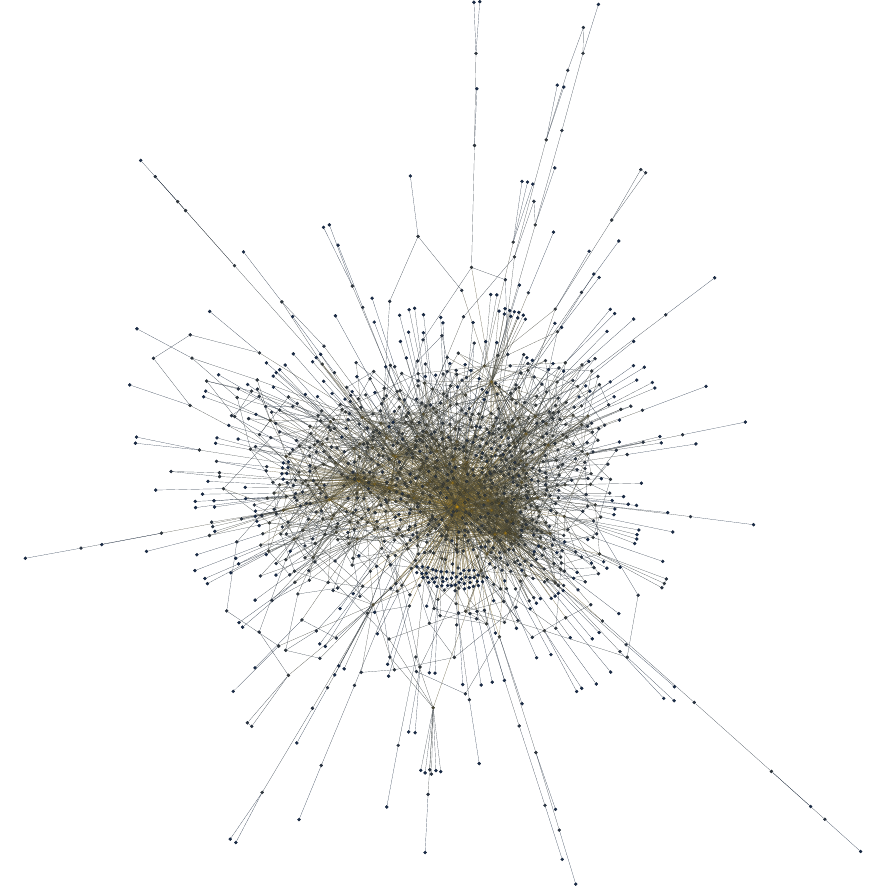}
		\caption*{(a) Largest Connected Component Visualization}
	\end{minipage}%
	\hfill
	\begin{minipage}[b]{0.42\linewidth}
		\centering
		\vspace{-1em} 
		\resizebox{\linewidth}{!}{%
			\begin{tabular}{lr}
				\toprule
				\textbf{Metric} & \textbf{Value} \\
				\midrule
				Number of nodes & 1293 \\
				Number of edges & 3592 \\
				Average degree & 5.56 \\
				Density & 0.0043 \\
				Is directed & False \\
				Maximum degree & 425 \\
				Minimum degree & 1 \\
				Average clustering coefficient & 0.1986 \\
				Number of triangles & 2236 \\
				Max degree centrality & 0.3289 \\
				Max betweenness centrality & 0.6172 \\
				Max closeness centrality & 0.4872 \\
				Diameter & 12 \\
				Radius & 6 \\
				Average shortest path length & 3.7671 \\
				Degree assortativity coefficient & -0.1036 \\
				\bottomrule
			\end{tabular}
		}
		\caption*{(b) Largest Connected Component Properties}
	\end{minipage}
\end{figure}

While the overall network reveals a fragmented periphery, the Largest Connected Component (LCC) serves as the true intellectual engine of digital network research. Visualized in Figure \ref{fig:Fig.9} using the force-directed Yifan Hu layout \cite{hu_visualizing_2015}, this dominant cluster comprises 1,293 nodes and 3,592 edges. It represents the most deeply interconnected ecosystem of publications, successfully filtering out isolated anomalies to highlight the works driving the field's theoretical evolution. The average degree of 5.56 indicates a healthy balance of academic discourse, where the typical publication engages with multiple theoretical perspectives simultaneously. 

Crucially, the LCC exhibits textbook "small-world" properties. With an incredibly short average path length of 3.7671 and a diameter of just 12, the network structure ensures that theoretical innovations can traverse the entire discipline rapidly. When paired with a relatively high average clustering coefficient (0.1986) and the presence of 2,236 closed triangles, this topology indicates that the field is composed of tightly knit research communities--such as distinct clusters focusing on organizational structures or digital media--that remain efficiently bridged to one another. Despite the low overall density (0.0043), this small-world architecture fosters the rapid cross-fertilization of ideas while maintaining rigorous, localized sub-disciplines.

However, the structural metrics also expose severe hierarchical power dynamics governing knowledge production. The degree distribution of the LCC is highly skewed; while the average degree is only 5.56, the maximum degree reaches an extraordinary 425.  This disparity confirms a near scale-free structure driven by preferential attachment. A concentrated fraction of foundational "hubs" dominate the network, capturing the vast majority of academic citations.

This top-down, hierarchical topology is further corroborated by the negative degree assortativity coefficient (-0.1036). In this network, highly cited canonical works are persistently linked to low-degree, newly published nodes. Rather than distinct research groups operating independently, emerging studies are structurally compelled to tether their novel inquiries to these massive, established hubs to gain academic legitimacy. This ensures cumulative academic progress but simultaneously creates theoretical bottlenecks, as reflected by the extreme maximum betweenness centrality (0.6172). To uncover exactly which foundational theories serve as these critical bottlenecks and bridges, the subsequent section presents a node-level centrality analysis of the LCC. 

\begin{figure}[htbp]
	\centering
	\caption{\label{fig:Fig.10}Comparison to the BA Model}
	\includegraphics[width = 0.5\linewidth]{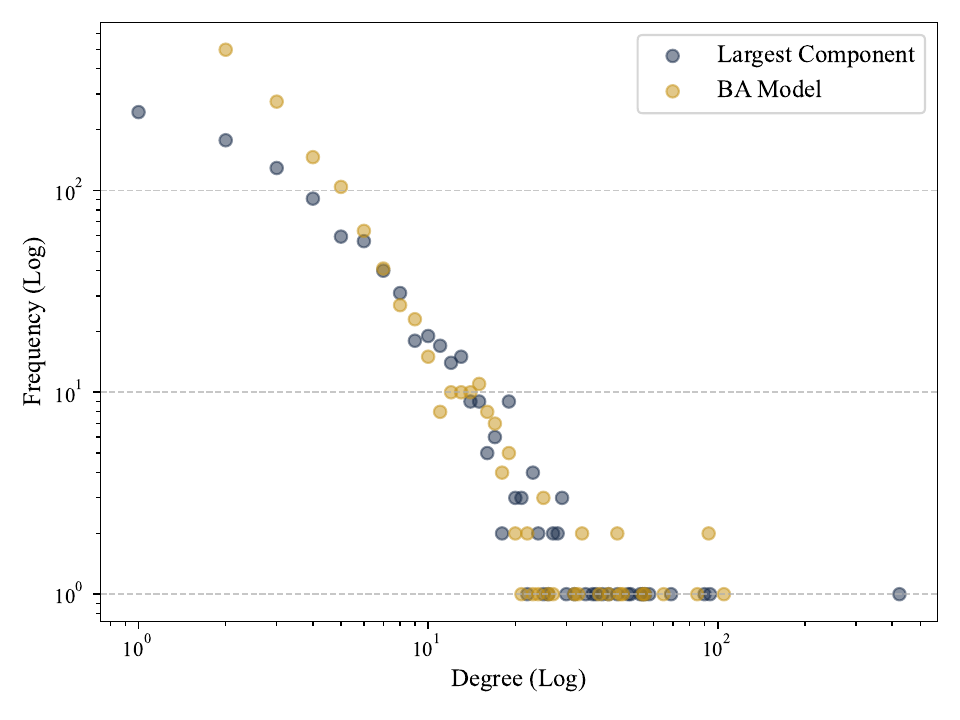}
\end{figure}

To understand the hierarchical topology of our citation network, we compared the empirical properties of the largest connected component (LCC) against the theoretical Barabási-Albert (BA) model (see Figure \ref{fig:Fig.10}). The BA model generates scale-free networks through the mechanism of preferential attachment, where a network is considered scale-free if its degree distribution follows a power law. Mathematically, the probability $P(k)$ that a node interacts with $k$ other nodes (in our case, accumulates citations) is expressed as:
\begin{equation}
	P(k) \sim k^{-\gamma}
\end{equation}
where $\gamma$ is a degree exponent typically falling in the range $2 < \gamma < 3$. 

As shown in Figure \ref{fig:Fig.10}, the empirical degree distribution of the LCC strongly aligns with these theoretical predictions, confirming the presence of massive intellectual hubs.  This mathematical structure captures the "rich-get-richer" dynamics of academic publishing: new papers preferentially attach to (cite) already heavily cited foundational works, driving the network's hierarchical topology. However, minor deviations from the idealized BA model--such as slightly lower clustering in the empirical LCC--suggest that citation behavior is not driven by preferential attachment alone, but is also moderated by domain-specific factors like collaborative preferences and thematic affinities.

Crucially, this scale-free topology carries significant sociological implications for the trajectory of the field. The structure reflects a pronounced 'Matthew Effect' in academic publishing, where a concentrated group of foundational theories disproportionately attracts new citations. Consequently, the presence of these massive hubs suggests a potential constraint on the field's intellectual diversity. As emerging research on novel digital phenomena—such as decentralized architectures or AI-driven filter bubbles—enters the literature, it is often forced to tether itself to these older, traditional hubs to gain academic legitimacy. While this ensures theoretical continuity, it raises the critical question of whether the overarching dominance of these traditional hubs inadvertently stifles the development of radically new frameworks needed to explain modern algorithmic complexities.

\subsubsection*{Centrality Analysis of Largest Connected Component}

The centrality analysis provides a detailed view of the most influential publications within the largest connected component of the citation network. By evaluating degree, betweenness, and closeness centrality, we can identify the pivotal nodes that shape the intellectual landscape and trace how foundational theories anchor modern digital research. 

\textbf{Degree centrality: The Intellectual Hubs} \\
Degree centrality identifies the most directly cited works in the network, reflecting their widespread recognition and foundational status. Degree centrality $C_D(v)$ measures the immediate influence of a paper/node $v$ based on direct citations:
\begin{equation}
	C_D(v) = \frac{\deg(v)}{N - 1}
\end{equation}
where $\deg(v)$ is the number of direct connections (citations) the node has, and $N$ is the total number of nodes in the network.The top-ranking works include:

\begin{itemize}
	\item \textit{"mcpherson (2001)": 0.3289}: This seminal work on homophily is cited by nearly one-third of the core network. Its exceptional prominence indicates that the "birds of a feather" principle remains the primary theoretical lens through which researchers understand digital network formation and segregation\cite{ WOS:000170748100017}.
	\item \textit{"centola (2010)": 0.0735}: Investigating how behaviors spread in online social networks, this study demonstrated the importance of clustered ties for sustained adoption, significantly influencing modern research on digital virality\cite{ WOS:000281485600033}.
	\item \textit{"snijders (2010)": 0.0704}: Focused on stochastic actor-oriented models, this paper serves as a critical methodological hub for researchers conducting dynamic network analyses\cite{ WOS:000274948700005}.
\end{itemize}

\textbf{Betweenness centrality: Bridging Disparate Subfields} \\ 
The betweenness centrality $C_B(v)$ for a node $v$ (representing a publication) is defined as:
\begin{equation}
	C_B(v) = \sum_{s \neq v \neq t} \frac{\sigma_{s,t}(v)}{\sigma_{s,t}}
\end{equation}
where $\sigma_{s,t}$ is the total number of shortest paths from node $s$ to node $t$, and $\sigma_{s,t}(v)$ is the number of those paths that pass through $v$. In the context of our citation network, a high $C_B(v)$ indicates that a paper frequently acts as a theoretical bridge between disparate research subfields.

\begin{itemize}
	\item \textit{"mcpherson (2001)": 0.6172}: In computational terms, this extraordinarily high score indicates that the paper frequently acts as the shortest path between disconnected nodes. Sociologically, it reveals that McPherson (2001)\cite{ WOS:000170748100017} is the critical bridge connecting traditional, offline sociological studies with modern, computational research on digital media. Researchers studying vastly different phenomena--from algorithmic echo chambers to political polarization--rely on this single anchor.
	\item \textit{"centola (2010)": 0.1011}: By emphasizing behavior diffusion, this work successfully links distinct research clusters focused on social influence, digital health, and communication networks.\cite{ WOS:000281485600033}.
	\item \textit{"burt (2000)": 0.0418}: By introducing the concept of "structural holes", this paper bridges classic theories of social capital with empirical applications in modern organizational and digital network studies.\cite{ WOS:000166194000008}.
\end{itemize}

Bridging nodes play a critical role in integrating interdisciplinary perspectives and enabling the exchange of ideas.

\textbf{Closeness centrality: The Theoretical Baseline} \\ 
Closeness centrality $C_C(v)$ measures the theoretical accessibility of a publication by calculating the inverse of the sum of its shortest path distances to all other nodes in the network:
\begin{equation}
	C_C(v) = \frac{N - 1}{\sum_{u \neq v} d(v, u)}
\end{equation}
where $d(v, u)$ is the shortest path distance (number of citation hops) between node $v$ and node $u$, and $N$ is the total number of nodes. In this study, works with high closeness centrality function as theoretical baselines, as they can be reached from almost any other paper in the network in very few citation steps. 

\begin{itemize}
	\item \textit{"mcpherson (2001)": 0.4872}: Its central location underscores its foundational role in shaping theoretical discourse\cite{ WOS:000170748100017}.
	\item \textit{"centola (2010)": 0.3896}: Acts as a unifying work that connects diverse methodological and substantive discussions\cite{ WOS:000281485600033}.
	\item \textit{"kossinets (2009)": 0.3818}: Examines temporal dynamics in social networks, contributing to understanding the evolution of online interactions\cite{Kossinets_2009}.
\end{itemize}

These works occupy positions of high theoretical accessibility. The overarching prominence of nodes like McPherson (2001) across all three metrics reflects their dual roles as both massive citation hubs and vital interdisciplinary bridges. Ultimately, these works ensure the theoretical stability of the field, providing the established frameworks necessary to investigate the volatile, algorithmic nature of modern digital contexts.

\subsection*{Keyword Co-occurrence Analysis}

\begin{figure}[htbp]
	\caption{\label{fig:Fig.3}Keywords Co-occurence Network and Its Density Layout}
	\includegraphics[width = 0.5\linewidth]{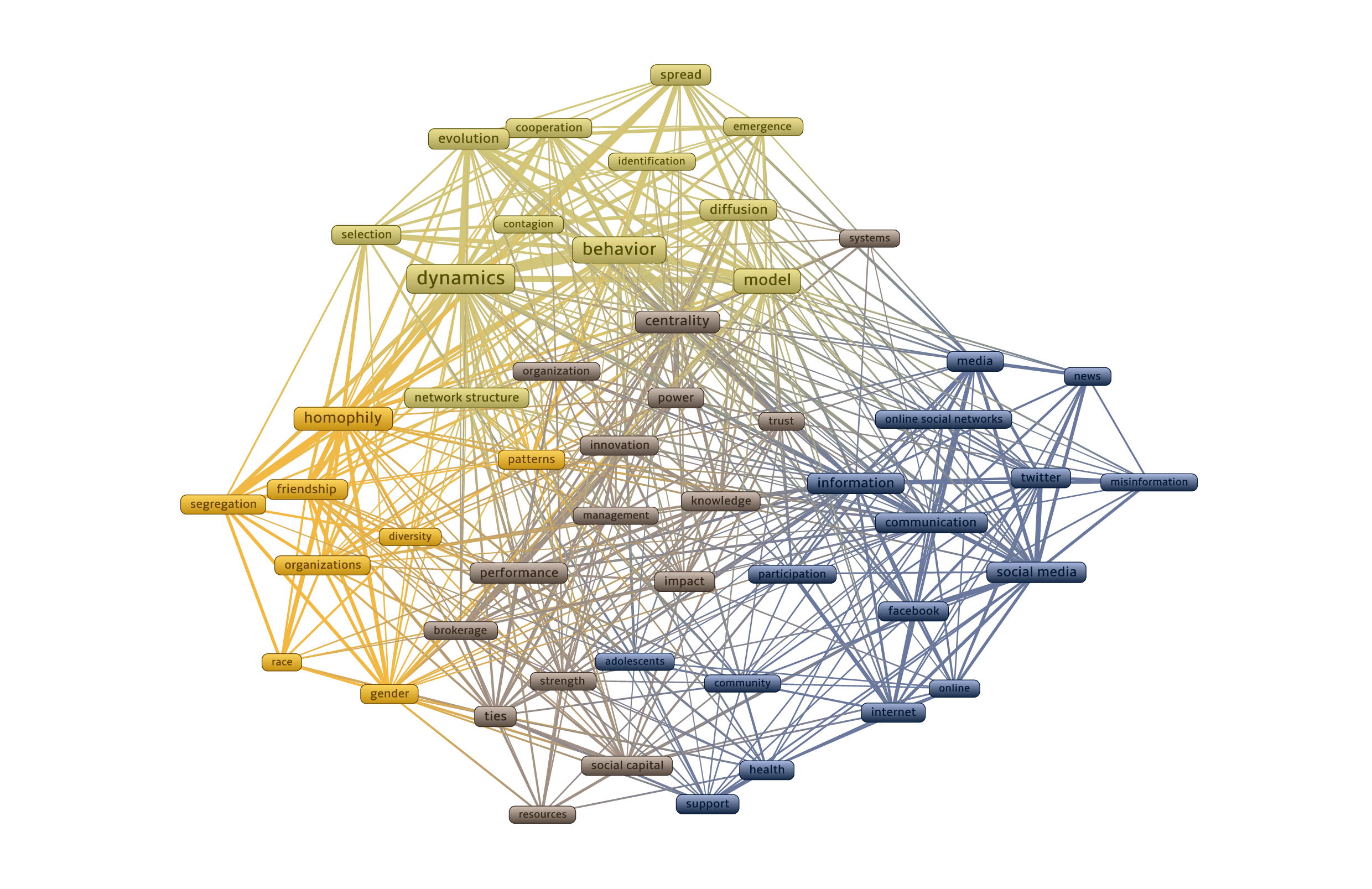}
	\includegraphics[width = 0.5\linewidth]{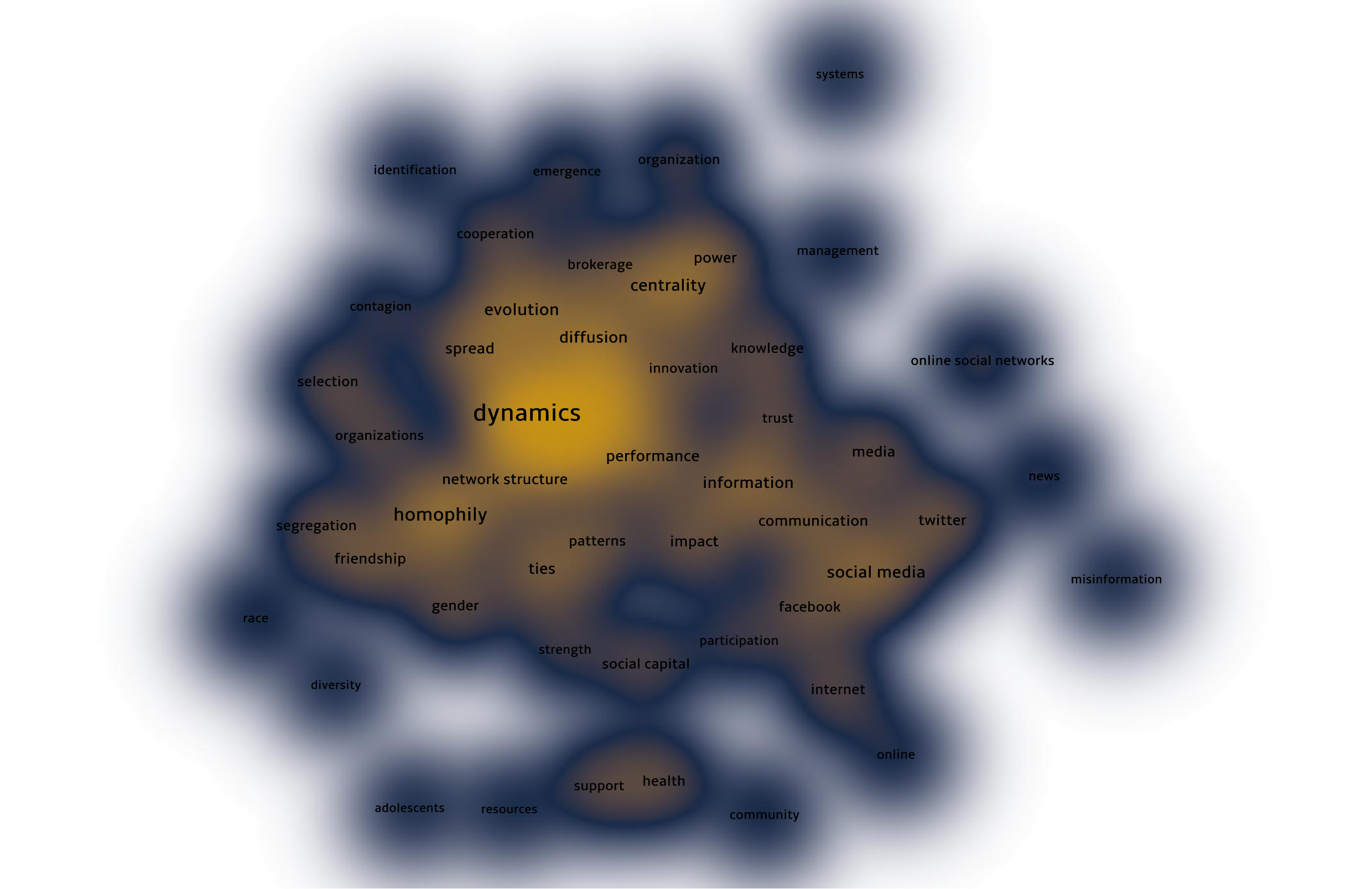}
\end{figure}

To move beyond the structural topology of the citation network and identify the dominant research frontiers, we conducted a keyword co-occurrence analysis using the full-counting method in VOSviewer. A minimum threshold of 35 co-occurrences was established, resulting in 60 keywords meeting the criterion. After meticulously excluding universal or purely generic search terms, the resulting semantic network is illustrated in Figure \ref{fig:Fig.3}. The maximum number of visualized lines was capped at 500 to optimize visual clarity and highlight only the most significant structural ties.

The network visualization highlights four distinct thematic clusters, capturing the intellectual evolution of digital network research from traditional sociological frameworks to modern algorithmic phenomena:
\begin{itemize}
	\item \textbf{Cluster 1: Social Dynamics and Tie Formation (Upper Left):} This cluster centers on foundational human behavior, featuring keywords such as \textit{dynamics}, \textit{behavior}, and \textit{homophily}. It emphasizes classic sociological processes like friendship formation and network segregation, confirming that traditional mechanisms of human connection remain vital for understanding digital interactions.
	\item \textbf{Cluster 2: Organizational Structures and Power (Brown):} Focusing on formal and institutional environments, this cluster features terms like \textit{organization}, \textit{power}, and \textit{centrality}. It reflects a sustained research interest in how digital networks alter governance, leadership, and knowledge transfer within bounded corporate or community structures.
	\item \textbf{Cluster 3: The Algorithmic Shift (Navy Blue):} This cluster explores the modern complexities of digital media communication. Dominated by keywords such as \textit{social media}, \textit{online social networks}, and \textit{misinformation}, it underscores a critical thematic pivot. Researchers are increasingly focusing on how algorithmic platforms actively mediate information dissemination, shifting the discipline's focus toward the societal impacts of networked communication and algorithmic bias.
	\item \textbf{Cluster 4: Methodological Bridges (Grey):} Centrally located, this cluster acts as the intellectual glue between the other distinct themes. It features common, bridging topics that connect traditional sociological theories with modern digital datasets, demonstrating the highly interdisciplinary nature of the field.
\end{itemize}

Complementing the network map, the density visualization (Figure \ref{fig:Fig.3}, right panel) provides a topological view of the field's "centers of gravity."  In this layout, areas illuminated in gold indicate high-frequency keyword concentrations, while blue areas denote less saturated topics. Rather than a fragmented discipline, the density map reveals highly concentrated, overlapping intellectual cores. Central terms such as \textit{dynamics} and \textit{information} bridge the traditional and modern clusters, highlighting that while the context of study has shifted toward digital platforms and misinformation, the core sociological mechanisms evaluating these networks remain tightly interwoven.

\subsection*{Citation Burst Analysis: Temporal Evolution and Foundational Anchors}

To move beyond the static semantic themes and map the temporal evolution of intellectual attention, we conducted a citation burst analysis. This allows us to identify paradigm shifts by detecting sudden spikes in specific citations over time. To achieve this, we transitioned our Web of Science (WoS) dataset into CiteSpace \cite{doi:10.1073/pnas.0307513100} to generate a GraphML file for advanced topological analysis in Gephi. 

Node selection was calibrated using the g-index \cite{egghe_theory_2006}, mathematically defined as $g^2 \leq \sum_{i \leq g} c_i, k \in \mathbb{Z}^{+}$, with the scaling factor $k$ set to 25 to ensure a comprehensive network structure. The temporal analysis was conducted using one-month intervals starting from 1982, and tie strengths were computed via the Jaccard index \cite{jaccard_distribution_1912}. 

\begin{figure}[htbp]
	\centering
	\caption{\label{fig:Fig.5}Top 25 References with the Strongest Citation Bursts}
	\includegraphics[width = \linewidth]{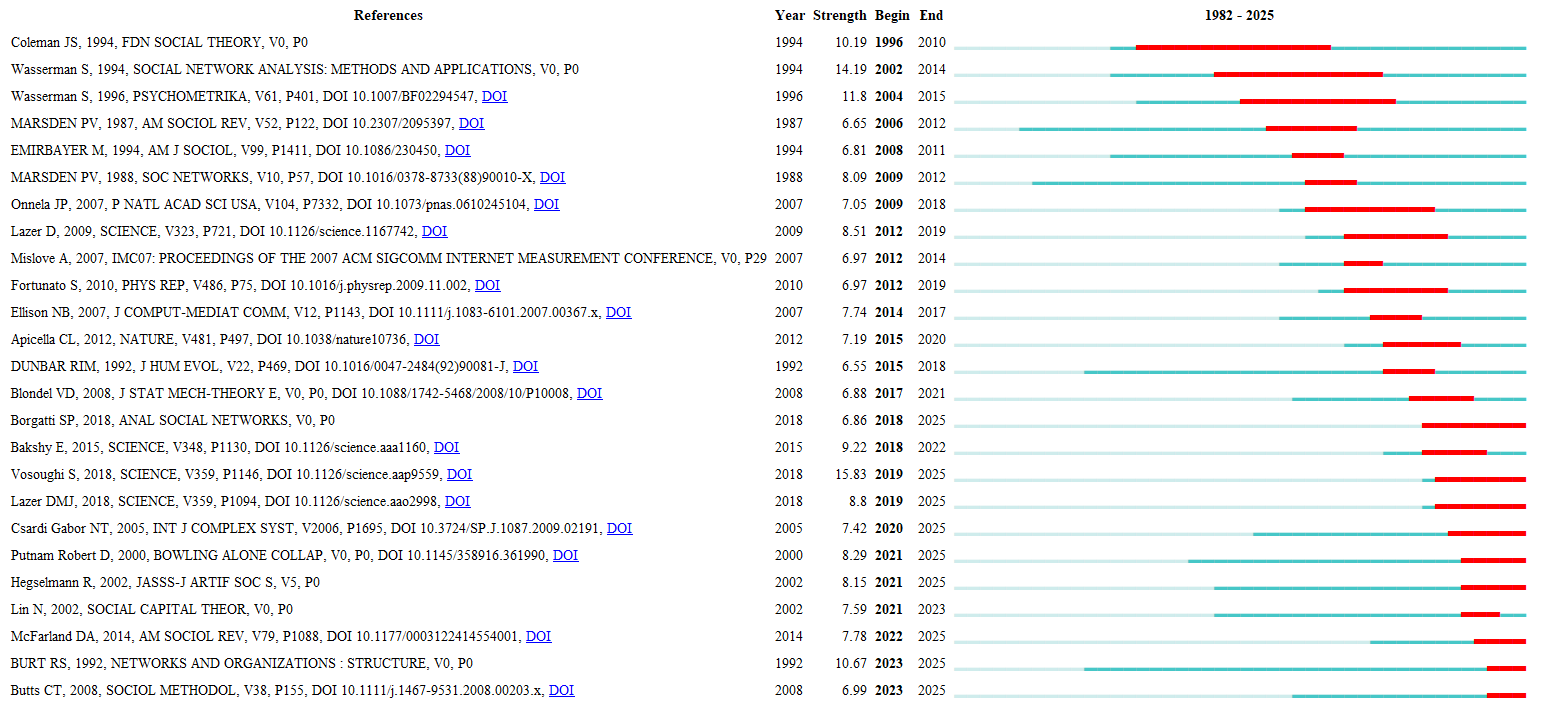}
\end{figure}

To identify historical periods of heightened academic focus, we applied the Kleinberg burst detection algorithm \cite{10.1145/775047.775061}. Figure \ref{fig:Fig.5} visualizes the top 25 references with the strongest citation bursts. The dark blue segments represent the entire lifespan of a publication's presence in the literature, while the red segments isolate the specific time intervals where the work experienced a statistically significant surge in citations, marking it as a critical historical catalyst for the field.

\begin{figure}[htbp]
	\centering
	\caption{\label{fig:Fig.7}Citation Network of Bursts}
	\includegraphics[width = 0.5\linewidth]{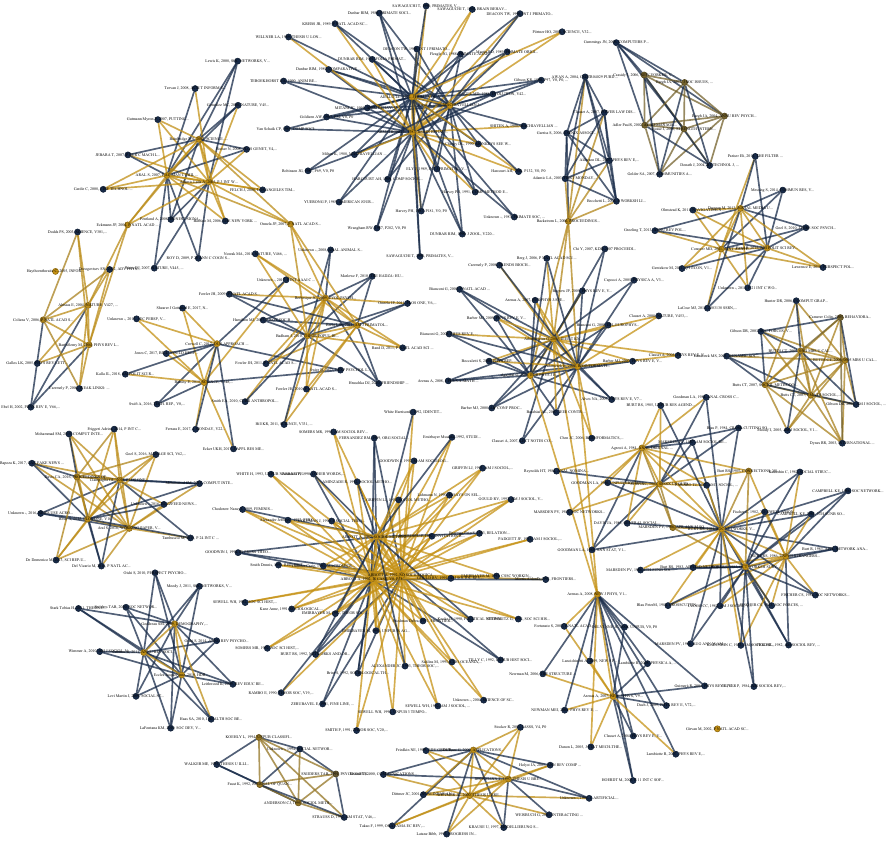}
\end{figure}

To understand how these burst publications interact structurally, we visualized their distinct citation network using the Fruchterman-Reingold layout \cite{fruchterman_graph_1991}, as illustrated in Figure \ref{fig:Fig.7}. Because each publication can only be cited once by a specific paper, the network ties are unweighted. High-degree nodes, acting as burst hubs, are highlighted in gold, while lower-degree nodes are depicted in navy blue. 

\begin{figure}[htbp]
	\centering
	\caption{\label{fig:Fig.8}Component 3 \& 4}
	\includegraphics[width = 0.45\linewidth]{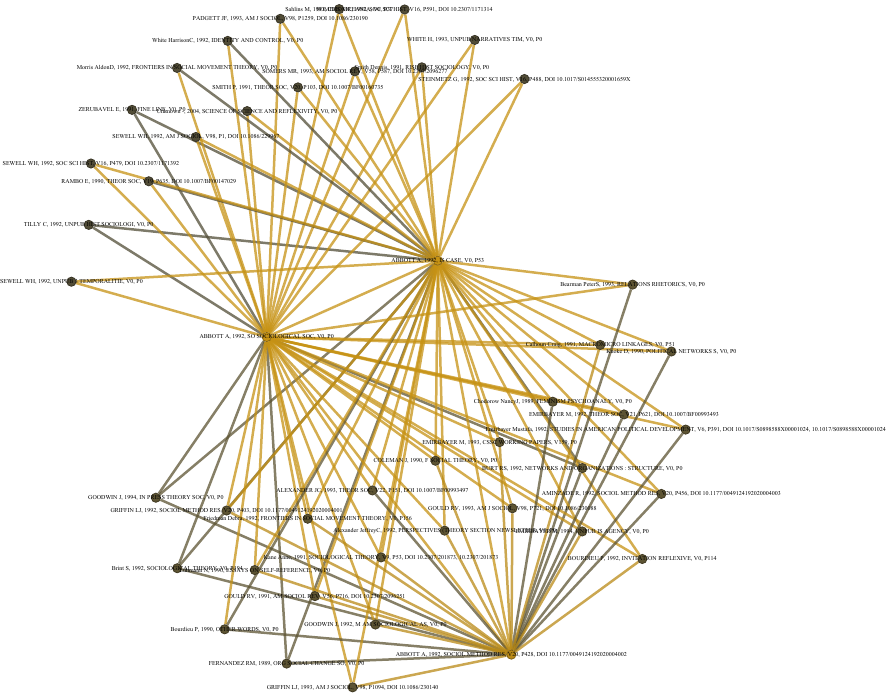}
	\includegraphics[width = 0.45\linewidth]{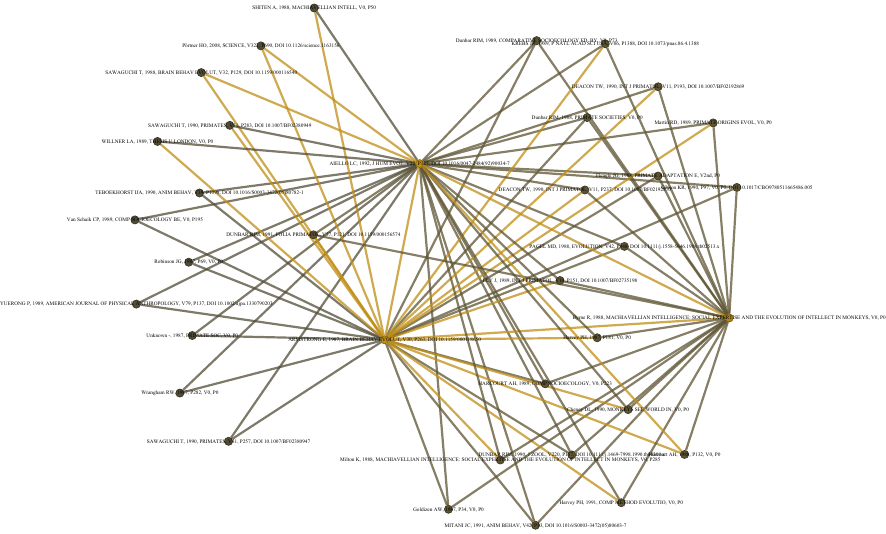}
\end{figure}

The resulting burst network is highly fragmented, consisting of 15 distinct connected components ranging in size from 1 to 46 nodes (incorporating isolated nodes resulting from the initial CiteSpace filtering process, as detailed in Appendix \ref*{app:config}). To examine this fragmentation, we extracted Components 3 and 4 as representative examples, mapped in Figure \ref{fig:Fig.8} using the ForceAtlas2 algorithm \cite{jacomy_forceatlas2_2014}.

A critical synthesis of these sub-components reveals a profound insight into the discipline's intellectual fact: the publications with the highest burst degrees do not actually originate from the specific field of digital social networks. Instead, they are foundational works drawn from classic sociology, physics, and mathematics. Rather than representing modern empirical studies on digital platforms, these burst nodes act as the indispensable theoretical and methodological cornerstones—the absolute baseline frameworks—upon which the entire modern study of algorithmic and digital networks was subsequently built.

\section{Information Flow and Inequality in Digital Networks}\label{Information Flow and Inequality in Digital Networks}

While digital networks possess the unprecedented capacity to rapidly disseminate information, they simultaneously embed and amplify profound structural inequalities. The dynamics of information flow represent a critical intersection of computational structure and sociological behavior. As evidenced by our bibliometric mapping, the literature has increasingly recognized that digital networks do not democratize influence; rather, they reshape disparities in access, visibility, and power. This section synthesizes the topological mechanisms that govern information flow, the algorithmic features that perpetuate inequality, and their broader implications for networked societies.

\subsection*{Topological Bottlenecks: Scale-Free Structures and Gatekeeping}

Information flow in digital networks is fundamentally constrained by their underlying topology. As demonstrated in our analysis of the network's core, digital communication infrastructures heavily exhibit scale-free properties, wherein a minute fraction of nodes accumulate a disproportionately large number of connections \cite{barabasi_emergence_1999}. Driven by the mechanism of preferential attachment, these highly connected hubs act as structural gatekeepers, facilitating rapid dissemination but also bottlenecking information access.  

Furthermore, while these networks display "small-world" properties that ensure short average path lengths \cite{watts_collective_1998}, this efficiency paradoxically exacerbates disparity. Because information must traverse the network via these massive central hubs, peripheral nodes remain marginalized in terms of both access and influence. As defined by foundational centrality theories \cite{freeman_centrality_1978}, this top-down, centralized hierarchy ensures that the power to frame academic, social, or political narratives remains concentrated in the hands of a few highly visible actors.

\subsection*{The Algorithmic Amplifier: Homophily and Echo Chambers}

Mechanisms of inequality in digital networks extend far beyond static topology; they are actively weaponized by the algorithmic systems integral to modern platforms. As highlighted by our keyword burst analysis, the field has aggressively shifted toward studying algorithmic effects and misinformation. Recommendation engines and ranking algorithms are inherently designed to prioritize engagement, creating a feedback loop that continually amplifies visibility for already-popular nodes \cite{lazer_rise_2015}. 

Crucially, these algorithms exploit the sociological principle of homophily—the tendency of individuals to connect with similar others \cite{WOS:000170748100017}. By exclusively recommending content that aligns with pre-existing user preferences and engagement patterns, algorithms structurally enforce network segregation.  This artificial clustering fosters impenetrable echo chambers and filter bubbles, severely limiting the diversity of information flow. Consequently, algorithmic bias not only creates a "rich-get-richer" dynamic for attention but also unintentionally marginalizes dissenting voices and underrepresented groups, fracturing the shared digital reality.

\subsection*{Systemic Barriers and the Digital Divide}

The topological and algorithmic inequalities within digital networks are compounded by offline, systemic barriers. The digital divide is not merely an artifact of network algorithms, but a reflection of unequal access to technological infrastructure and digital literacy. These structural barriers disproportionately affect economically disadvantaged populations, restricting their equitable participation in information networks. Ultimately, the synthesis of our bibliometric findings with these theoretical foundations reveals that digital social networks are not neutral conduits; they are complex socio-technical systems that, without deliberate intervention, inherently replicate and scale offline societal inequalities.

\subsection*{Empirical Insights: Cascades, Polarization, and Power Consolidation}

The structural and algorithmic inequalities inherent in digital networks generate profound empirical consequences, particularly concerning how information cascades through a population. As our centrality analysis demonstrated, networks are dominated by highly connected hubs that dictate both the breadth and directional flow of information \cite{centola_cascade_2007}.  While this centralization facilitates the rapid, widespread dissemination of viral content, it simultaneously entrenches inequality by granting dominant actors an asymmetric advantage over the narrative. 

Furthermore, this unequal dissemination is compounded by the phenomenon of network polarization. The high clustering coefficients observed in our topological analysis reflect the formation of highly segregated communities. Within these clusters, individuals primarily receive information that reinforces pre-existing beliefs, a dynamic that actively fosters polarization and echo chambers \cite{sunstein_law_2002, WOS:000170748100017}. This fragmentation inhibits constructive, cross-cutting dialogue, meaning that while flatter network structures might achieve more equitable dissemination, the current hub-dominated topology severely restricts exposure to diverse viewpoints. 

The societal consequences of these topological phenomena are profound. Digital concentration of influence inevitably translates into socio-economic power consolidation. Dominant nodes--whether they are monolithic tech corporations, political actors, or highly visible influencers--leverage their structural advantage within the network to extract disproportionate economic value and control public discourse, thereby deepening existing offline disparities.

\subsection*{Toward Equitable Networks through Structural and Algorithmic Interventions}

Addressing the deeply embedded inequalities of digital social networks requires a paradigm shift that moves beyond descriptive analysis toward active, multifaceted interventions. Because the inequalities are both structural and algorithmic, the solutions must operate on both levels simultaneously.

Structurally, the field must explore interventions that foster decentralized network architectures. Transitioning away from the strictly scale-free, hub-dominated topologies currently favored by commercial platforms can systematically reduce the bottlenecking of influence, promoting a more equitable distribution of access and visibility. [Image comparing centralized, decentralized, and distributed network architectures]

Algorithmically, developers and researchers must prioritize fairness over pure engagement metrics. Developing algorithms that consciously inject diversity into content dissemination and ensure equitable visibility can actively mitigate the harmful effects of preferential attachment \cite{binns_fairness_2018}. Balancing engagement-driven algorithms with fairness-focused designs is a critical step in dismantling artificially constructed filter bubbles.

Finally, these computational interventions must be grounded in real-world sociological efforts to enhance digital inclusion. Topologies and algorithms cannot be fixed if the fundamental users lack access. Policies aimed at expanding digital infrastructure and improving digital literacy are the foundational prerequisites to bridging systemic divides. Ultimately, mitigating inequality in digital networks demands robust public-private partnerships dedicated to democratizing both the access to the network and the algorithms that govern it.

\section{Information Propagation Dynamics}\label{Information Propagation Dynamics}

Information propagation dynamics examine how data, ideas, and influence cascade across networks. While traditional models focus on the speed and reach of dissemination, our analysis reveals that modern propagation is inextricably linked to the structural bottlenecks and behavioral echo chambers inherent in digital platforms. Understanding these processes is critical not only for analyzing communication efficiency but for addressing urgent societal challenges such as algorithmic misinformation and systemic polarization.

\subsection*{Structural and Behavioral Drivers of Propagation}

Network topology dictates the pathways of information flow. As highlighted by our topological findings, digital networks often feature pronounced core-periphery architectures \cite{borgatti_network_2011}.  Highly connected core nodes facilitate massive, efficient propagation, while peripheral nodes remain isolated, creating severe disparities in information access. In these hierarchical structures, overlapping communities must rely on boundary-spanning nodes to act as bridges. Without these high-betweenness connectors, information cannot traverse disconnected clusters, stifling the spread of diverse innovations \cite{rogers_diffusion_2003}.

However, structural topology is heavily moderated by individual behavioral psychology. Selective sharing—where users prioritize information that aligns with their pre-existing beliefs—artificially limits exposure to diverse perspectives and accelerates the formation of echo chambers \cite{WOS:000170748100017}. This homophilic behavior triggers reinforcement effects: repeated exposure to specific narratives within densely connected, insular clusters dramatically increases the likelihood of adoption, a mechanism that actively entrenches misinformation \cite{WOS:000281485600033}. Furthermore, the temporal nature of digital media dictates that emotionally charged content or breaking news exhibits extremely accelerated adoption curves followed by rapid decline, a volatility that favors sensationalism over factual dissemination \cite{WOS:000332845300020}. 

Ultimately, these propagation dynamics consistently reflect and reinforce offline inequalities. Peripheral users frequently function as ``invisible nodes,'' contributing minimally to the discourse \cite{zhao_effects_2011}. Conversely, ``visibility cascades'' occur when content introduced by central, highly visible actors spreads exponentially, allowing well-positioned hubs to monopolize the narrative \cite{borgatti_network_2011}. 

\section{Digital Networks and Micro-Macro Linkage}\label{Digital Networks and Micro-Macro Linkage}

To fully comprehend the inequalities of information propagation, social network analysis relies on the concept of micro-macro linkage: the dynamic interplay between localized, individual interactions (micro) and overarching systemic structures (macro). Traditional theories, such as Structural Balance Theory \cite{cartwright_structural_1956, heider_attitudes_1946}, posit that micro-level triadic relationships inevitably aggregate to dictate macro-level group divisions. 

\begin{figure}[htbp]
		\centering
		\caption{\label{fig:Fig.6}A social network perspective on Coleman’s micro–macro model}
		\begin{tikzpicture}
			\node (Macro-conditions) at (-5, 3) [place] {Macro-conditions};
			\node (Macro-outcomes)     at (5, 3) [place] {Macro-outcomes};
			\node (Micro-conditions)      at (-2.5,-1) [place] {Micro-conditions};
			\node (Micro-outcomes) at (2.5,-1) [place] {Micro-outcomes};
			\path[draw][-Stealth,dashed] (Macro-conditions) -- (Macro-outcomes);
			\path[draw][-Stealth] (Micro-conditions) -- (Micro-outcomes);
			\path[draw][-Stealth] (Macro-conditions) -- (Micro-conditions);
			\path[draw][-Stealth] (Micro-outcomes) -- (Macro-outcomes);
		\end{tikzpicture}
	\end{figure}
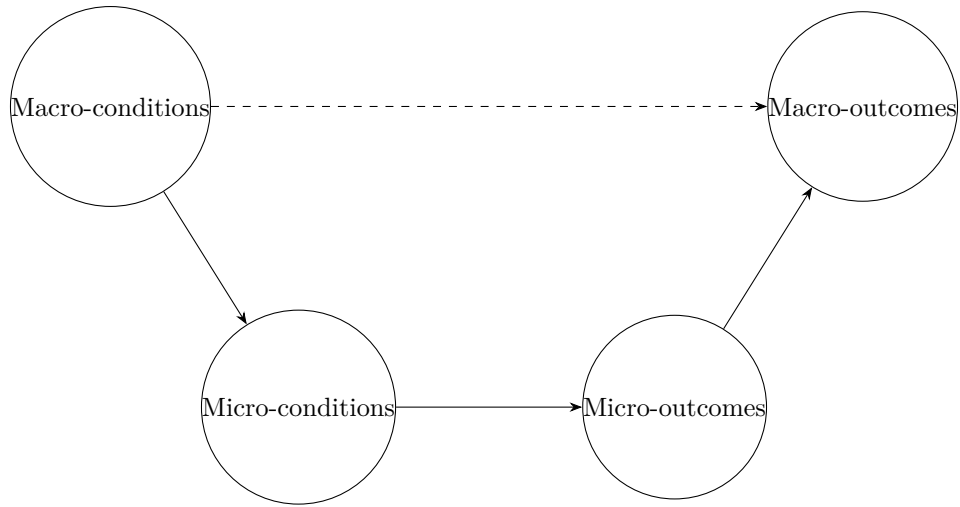

This reciprocal relationship is best illustrated by Coleman’s foundational micro-macro model \cite{coleman_foundations_1990}, conceptualized for network analysis by Raub et al.\ \cite{raub_micro-macro_2011} (see Figure \ref{fig:Fig.6}).
	
	The model maps a continuous feedback loop: bottom-up interactions (e.g., individual friendship ties) aggregate to form macro-level structures (e.g., network clustering), while top-down constraints (e.g., network density) dictate the future opportunities available to the individual.

\subsection*{The Algorithmic Loop in Digital Contexts}

In the context of modern digital networks, this classic micro-macro linkage has been fundamentally disrupted and accelerated by algorithmic mediation. The scalability of digital platforms amplifies bottom-up dynamics to an unprecedented degree. Individual micro-behaviors—such as liking, sharing, or utilizing a hashtag—can aggregate instantly into macro-level global movements. For instance, the rapid adoption of the \#MeToo hashtag transformed isolated personal narratives into a coordinated, decentralized macro-structure \cite{manikonda_metoo_2018}. Similarly, the micro-level decision to retweet controversial content aggregates to fuel massive misinformation cascades \cite{vosoughi_spread_2018}.

Simultaneously, the top-down forces in Coleman’s model are now predominantly governed by proprietary platform algorithms.  Algorithms dictate the ``Macro-conditions'' by curating news feeds, shaping visibility, and enforcing personalization \cite{tomlein_audit_2021}. When algorithms continuously recommend content that aligns with a user's micro-level preferences, they artificially enforce homophily \cite{pariser_filter_2011}. This generates a vicious cycle: algorithmic curation (macro) restricts individual exposure to diverse views (micro), prompting users to engage only with polarized content (micro), which the algorithm then scales into impenetrable, society-wide echo chambers (macro) \cite{bakshy_exposure_2015, lazer_science_2018}. 

Understanding this digitally mediated micro-macro loop is paramount. It necessitates a multidisciplinary approach—combining sociology, computer science, and behavioral economics—to develop theoretical frameworks capable of untangling how algorithmic systems hijack traditional human network behaviors \cite{chowdhury_understanding_2023}.

\section{Conclusion and Future Directions}\label{Conclusion}

This paper has explored the multifaceted, evolutionary dynamics of digital social networks. By employing comprehensive bibliometric and topological analyses, we mapped the intellectual core of the discipline, identifying a scale-free, small-world architecture dominated by foundational theoretical hubs. Furthermore, our keyword and citation burst analyses revealed a critical historical paradigm shift: the field has decisively transitioned from mapping static, offline topological structures to investigating the dynamic, algorithmic mediation of modern communication. 

The synthesis of these computational findings with sociological theory yields a stark reality: digital platforms do not inherently democratize information. Instead, driven by preferential attachment and algorithmic engagement models, they frequently exacerbate structural inequalities, bottleneck influence, and foster dangerous polarization. 

The rapid pace of technological development continually outpaces traditional sociological frameworks. Consequently, rather than offering definitive finality, this work serves as an empirical foundation for urgently needed future investigations. The algorithmic mediation of the micro-macro linkage remains a field ripe for critical inquiry. Future research must prioritize interdisciplinary collaboration to design equitable, decentralized network architectures and fairness-driven algorithms. Ultimately, the study of digital networks is not merely a retrospective scholarly endeavor, but an ongoing, essential dialogue to ensure the equitable future of our digitally mediated society.

This paper has sought to explore the multifaceted dynamics of digital social networks, from their theoretical foundations to their practical implications. By examining the interplay between micro-level interactions and macro-level structures, we have shed light on the unique challenges and opportunities presented by digital platforms. The bibliometric analysis and discussion of key academic questions offer a snapshot of the evolving intellectual landscape, underscoring the depth and breadth of scholarly engagement with these issues.

Yet, much remains to be understood. The rapid pace of technological development continually reshapes the ways in which individuals connect, interact, and organize, often outpacing our theoretical frameworks. Algorithmic mediation, for instance, is a field ripe for further exploration, particularly in understanding its long-term sociological consequences. Similarly, the role of digital networks in exacerbating or mitigating inequalities warrants sustained and critical inquiry.

Rather than offering definitive conclusions, this work aims to serve as a foundation for future investigations, inviting researchers to deepen their engagement with the questions posed here. By fostering interdisciplinary collaboration and leveraging innovative methodologies, the academic community can continue to advance our understanding of digital networks and their impact on society.

In this sense, the study of digital networks is not merely a scholarly endeavor but an ongoing dialogue—a collective effort to grapple with the complexities of a digitally mediated world. This paper represents one step in that journey, with the hope that it will inspire further exploration and debate.

\section*{Declarations}
\textbf{Disclosure of AI Use:} The authors utilized Large Language Models (LLMs) for linguistic refinement and the optimization of visualization scripts. The final analysis and theoretical interpretations were performed entirely by the human authors.

\pagebreak
\appendix
\section{WoS Query}
\label{app:query}

The search strategy employed targeted research on social networks using the following query: 

{\ttfamily\NavyBlue
	TS=("social network*") AND \\
	TS=("digital network*" OR "online social network*" OR "virtual network*" OR \\
	"computer-mediated communication" OR "internet-based interaction*" OR \\    
	"social network theor*" OR "network dynamic*" OR "tie formation" OR \\
	"micro-macro linkage*" OR "relational dynamic*" OR \\
	"information flow" OR "digital inequality" OR "algorithm* intervention*" OR \\
	"structural position*" OR "network structure*" OR \\
	"information diffusion" OR "misinformation" OR "echo chamber*" OR \\
	"filter bubble*" OR "homophily" OR "brokerage") and Multidisciplinary Sciences or \\
	Sociology or Social Sciences Interdisciplinary or Social Sciences Mathematical \\
	Methods (Web of Science Categories) and Social Sciences Citation Index (SSCI) \\ 
	or Science Citation Index Expanded (SCI-EXPANDED) (Web of Science Index) and \\
	Multidisciplinary Sciences or Sociology or Social Sciences Interdisciplinary or \\
	Social Sciences Mathematical Methods or Mathematics Interdisciplinary \\
	Applications or Communication or Computer Science Interdisciplinary Applications \\
	(Web of Science Categories)
}

\section{CiteSpace Configuration}
\label{app:config}
{\ttfamily\NavyBlue
	\begin{verbatim}
			"NounPhraseWordsMax": 4,
			"Description": "Bibliometrics of Digital Social Networks",
			"MaximumGMLNodeLabelLength": 8,
			"FilterByIntrinsicCitations": true,
			"AuthorED": false,
			"MaxLinksToRetain": 2.5,
			"AuthorGP": false,
			"SaveMergedSliceFiles": false,
			"UseC3": true,
			"DimensionsEndpoint": "https://app.dimensions.ai/",
			"ExclusionList": true,
			"ExportMatrix": false,
			"minimumThresholdValue": 1,
			"AliasList": true,
			"PercentageNodesToLabel": 1,
			"BurstWeight": 0,
			"UseAuthorFullname": true,
			"ConceptTreeHome": "C:\\Users\\Amethystium\\.citespace",
			"LookBackTime": -1,
			"MaxLinksPerNode": 10,
			"NormalizeCitations": false,
			"ExportSpace": false,
			"NounPhraseWordsMin": 2,
			"GlobalCheck": false,
			"NodeDegreeWeighted": true,
			"JDIC": true,
			"ExportAbstract": false,
			"LinkWeightModificationRate": 1.0E-11
	\end{verbatim}
}

\bibliographystyle{apalike}
\bibliography{citation.bib}

@article{ WOS:000170748100017,
	Author = {McPherson, M and Smith-Lovin, L and Cook, JM},
	Title = {Birds of a feather: Homophily in social networks},
	Journal = {ANNUAL REVIEW OF SOCIOLOGY},
	Year = {2001},
	Volume = {27},
	Pages = {415-444},
	DOI = {10.1146/annurev.soc.27.1.415},
	ISSN = {0360-0572},
	Unique-ID = {WOS:000170748100017},
}

@incollection{ WOS:000166194000008,
	Author = {Burt, RS},
	Editor = {Staw, BM and Sutton, RI},
	Title = {The network structure of social capital},
	Booktitle = {RESEARCH IN ORGANIZATIONAL BEHAVIOR, VOL 22, 2000: AN ANNUAL SERIES OF
	ANALYTICAL ESSAYS AND CRITICAL REVIEWS},
	Series = {Research in Organizational Behavior},
	Year = {2000},
	Volume = {22},
	Pages = {345-423},
	DOI = {10.1016/S0191-3085(00)22009-1},
	ISSN = {0191-3085},
	ISBN = {0-7623-0641-6},
	ResearcherID-Numbers = {Burt, Ronald S./KHV-2991-2024},
	Unique-ID = {WOS:000166194000008},
}

@article{ WOS:000337300100030,
	Author = {Kramer, Adam D. I. and Guillory, Jamie E. and Hancock, Jeffrey T.},
	Title = {Experimental evidence of massive-scale emotional contagion through
	social networks},
	Journal = {PROCEEDINGS OF THE NATIONAL ACADEMY OF SCIENCES OF THE UNITED STATES OF
	AMERICA},
	Year = {2014},
	Volume = {111},
	Number = {24},
	Pages = {8788-8790},
	Month = {JUN 17},
	DOI = {10.1073/pnas.1320040111},
	ISSN = {0027-8424},
	ORCID-Numbers = {Hancock, Jeffrey/0000-0001-5367-2677},
	Unique-ID = {WOS:000337300100030},
}

@article{ WOS:000281485600033,
	Author = {Centola, Damon},
	Title = {The Spread of Behavior in an Online Social Network Experiment},
	Journal = {SCIENCE},
	Year = {2010},
	Volume = {329},
	Number = {5996},
	Pages = {1194-1197},
	Month = {SEP 3},
	DOI = {10.1126/science.1185231},
	ISSN = {0036-8075},
	EISSN = {1095-9203},
	Unique-ID = {WOS:000281485600033},
}

@article{ WOS:000274948700005,
	Author = {Snijders, Tom A. B. and van de Bunt, Gerhard G. and Steglich, Christian
	E. G.},
	Title = {Introduction to stochastic actor-based models for network dynamics},
	Journal = {SOCIAL NETWORKS},
	Year = {2010},
	Volume = {32},
	Number = {1, SI},
	Pages = {44-60},
	Month = {JAN},
	DOI = {10.1016/j.socnet.2009.02.004},
	ISSN = {0378-8733},
	EISSN = {1879-2111},
	ResearcherID-Numbers = {van de Bunt, Gerhard/F-5485-2013
	Snijders, Tom/F-2896-2012},
	ORCID-Numbers = {Steglich, Christian/0000-0002-9097-0873
	Snijders, Tom/0000-0003-3157-4157},
	Unique-ID = {WOS:000274948700005},
}

@article{ WOS:000269632400036,
	Author = {Eagle, Nathan and Pentland, Alex (Sandy) and Lazer, David},
	Title = {Inferring friendship network structure by using mobile phone data},
	Journal = {PROCEEDINGS OF THE NATIONAL ACADEMY OF SCIENCES OF THE UNITED STATES OF
	AMERICA},
	Year = {2009},
	Volume = {106},
	Number = {36},
	Pages = {15274-15278},
	Month = {SEP 8},
	DOI = {10.1073/pnas.0900282106},
	ISSN = {0027-8424},
	Unique-ID = {WOS:000269632400036},
}

@article{Kossinets_2009,
	Author = {Kossinets, Gueorgi and Watts, Duncan J.},
	Title = {Origins of Homophily in an Evolving Social Network},
	Journal = {AMERICAN JOURNAL OF SOCIOLOGY},
	Year = {2009},
	Volume = {115},
	Number = {2},
	Pages = {405-450},
	Month = {SEP},
	DOI = {10.1086/599247},
	ISSN = {0002-9602},
	EISSN = {1537-5390},
	ResearcherID-Numbers = {Watts, Duncan/J-6483-2012},
	Unique-ID = {WOS:000270921400003},
}

@article{ WOS:000332845300020,
	Author = {Coviello, Lorenzo and Sohn, Yunkyu and Kramer, Adam D. I. and Marlow,
	Cameron and Franceschetti, Massimo and Christakis, Nicholas A. and
	Fowler, James H.},
	Title = {Detecting Emotional Contagion in Massive Social Networks},
	Journal = {PLOS ONE},
	Year = {2014},
	Volume = {9},
	Number = {3},
	Month = {MAR 12},
	DOI = {10.1371/journal.pone.0090315},
	Article-Number = {e90315},
	ISSN = {1932-6203},
	ResearcherID-Numbers = {Christakis, Nicholas/B-6690-2008
	Fowler, James/C-2750-2008},
	ORCID-Numbers = {Fowler, James/0000-0001-7795-1638},
	Unique-ID = {WOS:000332845300020},
}

@article{ WOS:A1982NM27900003,
	Author = {FRIEDKIN, NE},
	Title = {INFORMATION-FLOW THROUGH STRONG AND WEAK TIES IN INTRA-ORGANIZATIONAL
	SOCIAL NETWORKS},
	Journal = {SOCIAL NETWORKS},
	Year = {1982},
	Volume = {3},
	Number = {4},
	Pages = {273-285},
	DOI = {10.1016/0378-8733(82)90003-X},
	ISSN = {0378-8733},
	ResearcherID-Numbers = {Friedkin, Noah/B-5135-2009},
	Unique-ID = {WOS:A1982NM27900003},
}

@article{ WOS:000209041300003,
	Author = {Bruns, Axel},
	Title = {HOW LONG IS A TWEET? MAPPING DYNAMIC CONVERSATION NETWORKS ON
	TWITTER USING GAWK AND GEPHI},
	Journal = {INFORMATION COMMUNICATION \& SOCIETY},
	Year = {2012},
	Volume = {15},
	Number = {9},
	Pages = {1323-1351},
	DOI = {10.1080/1369118X.2011.635214},
	ISSN = {1369-118X},
	EISSN = {1468-4462},
	ResearcherID-Numbers = {Bruns, Axel/I-9877-2012
	},
	ORCID-Numbers = {Bruns, Axel/0000-0002-3943-133X},
	Unique-ID = {WOS:000209041300003},
}

@article{ WOS:000352789600005,
	Author = {Hargittai, Eszter},
	Title = {Is Bigger Always Better? Potential Biases of Big Data Derived from
	Social Network Sites},
	Journal = {ANNALS OF THE AMERICAN ACADEMY OF POLITICAL AND SOCIAL SCIENCE},
	Year = {2015},
	Volume = {659},
	Number = {1},
	Pages = {63-76},
	Month = {MAY},
	DOI = {10.1177/0002716215570866},
	ISSN = {0002-7162},
	EISSN = {1552-3349},
	ORCID-Numbers = {Hargittai, Eszter/0000-0003-4199-4868},
	Unique-ID = {WOS:000352789600005},
}

@article{ WOS:000274745000002,
	Author = {West, Anne and Lewis, Jane and Currie, Peter},
	Title = {Students' Facebook `friends': public and private spheres},
	Journal = {JOURNAL OF YOUTH STUDIES},
	Year = {2009},
	Volume = {12},
	Number = {6},
	Pages = {615-627},
	DOI = {10.1080/13676260902960752},
	Article-Number = {PII 915909652},
	ISSN = {1367-6261},
	EISSN = {1469-9680},
	ORCID-Numbers = {West, Anne/0000-0003-2932-7667},
	Unique-ID = {WOS:000274745000002},
}

@article{ WOS:000334339000015,
	Author = {Garcia-Herranz, Manuel and Moro, Esteban and Cebrian, Manuel and
	Christakis, Nicholas A. and Fowler, James H.},
	Title = {Using Friends as Sensors to Detect Global-Scale Contagious Outbreaks},
	Journal = {PLOS ONE},
	Year = {2014},
	Volume = {9},
	Number = {4},
	Month = {APR 9},
	DOI = {10.1371/journal.pone.0092413},
	Article-Number = {e92413},
	ISSN = {1932-6203},
	ResearcherID-Numbers = {Moro, Esteban/AAB-1159-2019
	Christakis, Nicholas/B-6690-2008
	Garcia-Herranz, Manuel/ABB-2617-2021
	Fowler, James/C-2750-2008
	Garcia-Herranz, Manuel/L-3213-2013
	},
	ORCID-Numbers = {Fowler, James/0000-0001-7795-1638
	Garcia-Herranz, Manuel/0000-0002-4252-4975
	Cebrian, Manuel/0000-0002-3681-7982
	MORO, ESTEBAN/0000-0003-2894-1024},
	Unique-ID = {WOS:000334339000015},
}

@article{ WOS:A1983QK14400006,
	Author = {DUNN, WN},
	Title = {SOCIAL NETWORK THEORY},
	Journal = {KNOWLEDGE-CREATION DIFFUSION UTILIZATION},
	Year = {1983},
	Volume = {4},
	Number = {3},
	Pages = {453-461},
	DOI = {10.1177/107554708300400306},
	ISSN = {0164-0259},
	Unique-ID = {WOS:A1983QK14400006},
}

@article{ WOS:A1982NG01900001,
	Author = {PERRUCCI, R and TARG, DB},
	Title = {NETWORK STRUCTURE AND REACTIONS TO PRIMARY DEVIANCE OF MENTAL-PATIENTS},
	Journal = {JOURNAL OF HEALTH AND SOCIAL BEHAVIOR},
	Year = {1982},
	Volume = {23},
	Number = {1},
	Pages = {2-17},
	DOI = {10.2307/2136385},
	ISSN = {0022-1465},
	EISSN = {2150-6000},
	Unique-ID = {WOS:A1982NG01900001},
}

@article{ WOS:A1969F015300009,
	Author = {PRITCHARD, A},
	Title = {STATISTICAL BIBLIOGRAPHY OR BIBLIOMETRICS},
	Journal = {JOURNAL OF DOCUMENTATION},
	Year = {1969},
	Volume = {25},
	Number = {4},
	Pages = {348+},
	ISSN = {0022-0418},
	EISSN = {1758-7379},
	Unique-ID = {WOS:A1969F015300009},
}

@article{ WOS:000375954300052,
	Author = {Yu, Dejian and Liao, Huchang},
	Title = {Visualization and quantitative research on intuitionistic fuzzy studies},
	Journal = {JOURNAL OF INTELLIGENT \& FUZZY SYSTEMS},
	Year = {2016},
	Volume = {30},
	Number = {6},
	Pages = {3653-3663},
	DOI = {10.3233/IFS-162111},
	ISSN = {1064-1246},
	EISSN = {1875-8967},
	ResearcherID-Numbers = {Dejian, Yu/IAM-0202-2023
	Liao, Huchang/F-9716-2015},
	ORCID-Numbers = {Yu, Dejian/0000-0003-2796-9148
	Liao, Huchang/0000-0001-8278-3384},
	Unique-ID = {WOS:000375954300052},
}

@article{ WOS:000278695500019,
	Author = {van Eck, Nees Jan and Waltman, Ludo},
	Title = {Software survey: VOSviewer, a computer program for bibliometric mapping},
	Journal = {SCIENTOMETRICS},
	Year = {2010},
	Volume = {84},
	Number = {2},
	Pages = {523-538},
	Month = {AUG},
	DOI = {10.1007/s11192-009-0146-3},
	ISSN = {0138-9130},
	EISSN = {1588-2861},
	ResearcherID-Numbers = {Waltman, Ludo/B-5561-2008
	van Eck, Nees Jan/B-6042-2008},
	ORCID-Numbers = {van Eck, Nees Jan/0000-0001-8448-4521},
	Unique-ID = {WOS:000278695500019},
}

@article{doi:10.1073/pnas.0307513100,
	author = {Chaomei Chen },
	title = {Searching for intellectual turning points: Progressive knowledge domain visualization},
	journal = {Proceedings of the National Academy of Sciences},
	volume = {101},
	number = {suppl\_1},
	pages = {5303-5310},
	year = {2004},
	doi = {10.1073/pnas.0307513100},
	URL = {https://www.pnas.org/doi/abs/10.1073/pnas.0307513100},
	eprint = {https://www.pnas.org/doi/pdf/10.1073/pnas.0307513100}}

@article{
		doi:10.1126/science.1167742,
		author = {David Lazer  and Alex Pentland  and Lada Adamic  and Sinan Aral  and Albert-László Barabási  and Devon Brewer  and Nicholas Christakis  and Noshir Contractor  and James Fowler  and Myron Gutmann  and Tony Jebara  and Gary King  and Michael Macy  and Deb Roy  and Marshall Van Alstyne },
		title = {Computational Social Science},
		journal = {Science},
		volume = {323},
		number = {5915},
		pages = {721-723},
		year = {2009},
		doi = {10.1126/science.1167742},
		URL = {https://www.science.org/doi/abs/10.1126/science.1167742},
		eprint = {https://www.science.org/doi/pdf/10.1126/science.1167742}}

@article{egghe_theory_2006,
		title = {Theory and practise of the g-index},
		volume = {69},
		issn = {1588-2861},
		url = {https://doi.org/10.1007/s11192-006-0144-7},
		doi = {10.1007/s11192-006-0144-7},
		language = {en},
		number = {1},
		urldate = {2024-11-26},
		journal = {Scientometrics},
		author = {Egghe, Leo},
		month = oct,
		year = {2006},
		pages = {131--152},
	}

@article{jaccard_distribution_1912,
		title = {The {Distribution} of the {Flora} in the {Alpine} {Zone}.},
		volume = {11},
		issn = {1469-8137},
		url = {https://onlinelibrary.wiley.com/doi/abs/10.1111/j.1469-8137.1912.tb05611.x},
		doi = {10.1111/j.1469-8137.1912.tb05611.x},
		language = {en},
		number = {2},
		urldate = {2024-11-26},
		journal = {New Phytologist},
		author = {Jaccard, Paul},
		year = {1912},
		note = {\_eprint: https://onlinelibrary.wiley.com/doi/pdf/10.1111/j.1469-8137.1912.tb05611.x},
		pages = {37--50},
	}

@inproceedings{10.1145/775047.775061,
		author = {Kleinberg, Jon},
		title = {Bursty and hierarchical structure in streams},
		year = {2002},
		isbn = {158113567X},
		publisher = {Association for Computing Machinery},
		address = {New York, NY, USA},
		url = {https://doi.org/10.1145/775047.775061},
		doi = {10.1145/775047.775061},
		booktitle = {Proceedings of the Eighth ACM SIGKDD International Conference on Knowledge Discovery and Data Mining},
		pages = {91–101},
		numpages = {11},
		location = {Edmonton, Alberta, Canada},
		series = {KDD '02}
	}

@article{raub_micro-macro_2011,
		title = {Micro-{Macro} {Links} and {Microfoundations} in {Sociology}},
		volume = {35},
		issn = {0022-250X},
		url = {https://doi.org/10.1080/0022250X.2010.532263},
		doi = {10.1080/0022250X.2010.532263},
		number = {1-3},
		urldate = {2024-11-27},
		journal = {The Journal of Mathematical Sociology},
		author = {RAUB, WERNER and BUSKENS, VINCENT and VAN ASSEN, MARCEL A. L. M.},
		month = jan,
		year = {2011},
		note = {Publisher: Routledge
		\_eprint: https://doi.org/10.1080/0022250X.2010.532263},
		pages = {1--25},
	}

@book{coleman_foundations_1990,
		title = {Foundations of social theory},
		isbn = {978-0-674-31225-8 978-0-674-31226-5},
		url = {http://archive.org/details/foundationsofsoc0000cole},
		language = {eng},
		urldate = {2024-11-27},
		publisher = {Cambridge, Mass. : Belknap Press of Harvard University Press},
		author = {Coleman, James S.},
		collaborator = {{Internet Archive}},
		year = {1990},
	}

@article{cartwright_structural_1956,
		title = {Structural balance: a generalization of {Heider}'s theory},
		volume = {63},
		issn = {1939-1471},
		shorttitle = {Structural balance},
		doi = {10.1037/h0046049},
		number = {5},
		journal = {Psychological Review},
		author = {Cartwright, Dorwin and Harary, Frank},
		year = {1956},
		note = {Place: US
		Publisher: American Psychological Association},
		pages = {277--293},
	}

@article{freeman_centrality_1978,
		title = {Centrality in social networks conceptual clarification},
		volume = {1},
		issn = {0378-8733},
		url = {https://www.sciencedirect.com/science/article/pii/0378873378900217},
		doi = {10.1016/0378-8733(78)90021-7},
		number = {3},
		urldate = {2024-11-27},
		journal = {Social Networks},
		author = {Freeman, Linton C.},
		month = jan,
		year = {1978},
		pages = {215--239},
	}

@article{barabasi_emergence_1999,
		title = {Emergence of {Scaling} in {Random} {Networks}},
		volume = {286},
		url = {https://www.science.org/doi/10.1126/science.286.5439.509},
		doi = {10.1126/science.286.5439.509},
		number = {5439},
		urldate = {2024-11-27},
		journal = {Science},
		author = {Barabási, Albert-László and Albert, Réka},
		month = oct,
		year = {1999},
		note = {Publisher: American Association for the Advancement of Science},
		pages = {509--512},
	}

@article{heider_attitudes_1946,
		title = {Attitudes and {Cognitive} {Organization}},
		volume = {21},
		issn = {0022-3980},
		url = {https://doi.org/10.1080/00223980.1946.9917275},
		doi = {10.1080/00223980.1946.9917275},
		number = {1},
		urldate = {2024-11-27},
		journal = {The Journal of Psychology},
		author = {Heider, Fritz},
		month = jan,
		year = {1946},
		pmid = {21010780},
		note = {Publisher: Routledge
		\_eprint: https://doi.org/10.1080/00223980.1946.9917275},
		pages = {107--112},
	}

@article{manikonda_metoo_2018,
		series = {Lecture {Notes} in {Computer} {Science} (including subseries {Lecture} {Notes} in {Artificial} {Intelligence} and {Lecture} {Notes} in {Bioinformatics})},
		title = {\#metoo through the lens of social media: 11th {International} {Conference} on {Social} {Computing}, {Behavioral}-{Cultural} {Modeling}, and {Prediction} conference and {Behavior} {Representation} in {Modeling} and {Simulation}, {SBP}-{BRiMS} 2018},
		issn = {9783319933719},
		shorttitle = {\#metoo through the lens of social media},
		url = {http://www.scopus.com/inward/record.url?scp=85049780527&partnerID=8YFLogxK},
		doi = {10.1007/978-3-319-93372-6_13},
		urldate = {2024-11-29},
		journal = {Social, Cultural, and Behavioral Modeling - 11th International Conference, SBP-BRiMS 2018, Proceedings},
		author = {Manikonda, Lydia and Beigi, Ghazaleh and Kambhampati, Subbarao and Liu, Huan},
		editor = {Bisgin, Halil and Thomson, Robert and Hyder, Ayaz and Dancy, Christopher},
		year = {2018},
		note = {Publisher: Springer Verlag},
		pages = {104--110},
	}

@article{vosoughi_spread_2018,
		title = {The spread of true and false news online},
		volume = {359},
		url = {https://www.science.org/doi/10.1126/science.aap9559},
		doi = {10.1126/science.aap9559},
		number = {6380},
		urldate = {2024-11-29},
		journal = {Science},
		author = {Vosoughi, Soroush and Roy, Deb and Aral, Sinan},
		month = mar,
		year = {2018},
		note = {Publisher: American Association for the Advancement of Science},
		pages = {1146--1151},
	}

@article{granovetter_strength_1973,
		title = {The {Strength} of {Weak} {Ties}},
		volume = {78},
		issn = {0002-9602},
		url = {https://www.journals.uchicago.edu/doi/10.1086/225469},
		doi = {10.1086/225469},
		number = {6},
		urldate = {2024-11-29},
		journal = {American Journal of Sociology},
		author = {Granovetter, Mark S.},
		month = may,
		year = {1973},
		note = {Publisher: The University of Chicago Press},
		pages = {1360--1380},
	}

@inproceedings{tomlein_audit_2021,
		address = {New York, NY, USA},
		series = {{RecSys} '21},
		title = {An {Audit} of {Misinformation} {Filter} {Bubbles} on {YouTube}: {Bubble} {Bursting} and {Recent} {Behavior} {Changes}},
		isbn = {978-1-4503-8458-2},
		shorttitle = {An {Audit} of {Misinformation} {Filter} {Bubbles} on {YouTube}},
		url = {https://dl.acm.org/doi/10.1145/3460231.3474241},
		doi = {10.1145/3460231.3474241},
		urldate = {2024-11-28},
		booktitle = {Proceedings of the 15th {ACM} {Conference} on {Recommender} {Systems}},
		publisher = {Association for Computing Machinery},
		author = {Tomlein, Matus and Pecher, Branislav and Simko, Jakub and Srba, Ivan and Moro, Robert and Stefancova, Elena and Kompan, Michal and Hrckova, Andrea and Podrouzek, Juraj and Bielikova, Maria},
		month = sep,
		year = {2021},
		pages = {1--11},
	}

@article{bakshy_exposure_2015,
		title = {Exposure to ideologically diverse news and opinion on {Facebook}},
		volume = {348},
		url = {https://www.science.org/doi/10.1126/science.aaa1160},
		doi = {10.1126/science.aaa1160},
		language = {en-US},
		number = {6239},
		urldate = {2024-11-29},
		journal = {Science},
		author = {Bakshy, Eytan and Messing, Solomon and Adamic, Lada A.},
		month = jun,
		year = {2015},
		note = {Publisher: American Association for the Advancement of Science},
		pages = {1130--1132},
	}

@book{pariser_filter_2011,
	title = {The {Filter} {Bubble}: {How} the {New} {Personalized} {Web} {Is} {Changing} {What} {We} {Read} and {How} {We} {Think}},
	isbn = {978-1-101-51512-9},
	shorttitle = {The {Filter} {Bubble}},
	language = {en},
	publisher = {Penguin},
	author = {Pariser, Eli},
	month = may,
	year = {2011},
	note = {Google-Books-ID: wcalrOI1YbQC},
}

@article{lazer_science_2018,
	title = {The science of fake news},
	volume = {359},
	url = {https://www.science.org/doi/10.1126/science.aao2998},
	doi = {10.1126/science.aao2998},
	number = {6380},
	urldate = {2024-11-29},
	journal = {Science},
	author = {Lazer, David M. J. and Baum, Matthew A. and Benkler, Yochai and Berinsky, Adam J. and Greenhill, Kelly M. and Menczer, Filippo and Metzger, Miriam J. and Nyhan, Brendan and Pennycook, Gordon and Rothschild, David and Schudson, Michael and Sloman, Steven A. and Sunstein, Cass R. and Thorson, Emily A. and Watts, Duncan J. and Zittrain, Jonathan L.},
	month = mar,
	year = {2018},
	note = {Publisher: American Association for the Advancement of Science},
	pages = {1094--1096},
}

@article{chowdhury_understanding_2023,
	title = {Understanding misinformation infodemic during public health emergencies due to large-scale disease outbreaks: a rapid review},
	volume = {31},
	issn = {1613-2238},
	shorttitle = {Understanding misinformation infodemic during public health emergencies due to large-scale disease outbreaks},
	url = {https://doi.org/10.1007/s10389-021-01565-3},
	doi = {10.1007/s10389-021-01565-3},
	language = {en},
	number = {4},
	urldate = {2024-11-29},
	journal = {Journal of Public Health},
	author = {Chowdhury, Nashit and Khalid, Ayisha and Turin, Tanvir C.},
	month = apr,
	year = {2023},
	pages = {553--573},
}

@article{watts_collective_1998,
	title = {Collective dynamics of ‘small-world’ networks},
	volume = {393},
	copyright = {1998 Macmillan Magazines Ltd.},
	issn = {1476-4687},
	url = {https://www.nature.com/articles/30918},
	doi = {10.1038/30918},
	language = {en},
	number = {6684},
	urldate = {2024-11-29},
	journal = {Nature},
	author = {Watts, Duncan J. and Strogatz, Steven H.},
	month = jun,
	year = {1998},
	note = {Publisher: Nature Publishing Group},
	pages = {440--442},
}

@article{lazer_rise_2015,
	title = {The rise of the social algorithm},
	volume = {348},
	url = {https://www.science.org/doi/10.1126/science.aab1422},
	doi = {10.1126/science.aab1422},
	language = {en-US},
	number = {6239},
	urldate = {2024-11-29},
	journal = {Science},
	author = {Lazer, David},
	month = jun,
	year = {2015},
	note = {Publisher: American Association for the Advancement of Science},
	pages = {1090--1091},
}

@article{centola_cascade_2007,
	title = {Cascade dynamics of complex propagation},
	volume = {374},
	issn = {0378-4371},
	url = {https://www.sciencedirect.com/science/article/pii/S0378437106007679},
	doi = {10.1016/j.physa.2006.06.018},
	number = {1},
	urldate = {2024-11-29},
	journal = {Physica A: Statistical Mechanics and its Applications},
	author = {Centola, Damon and Eguíluz, Víctor M. and Macy, Michael W.},
	month = jan,
	year = {2007},
	pages = {449--456},
}

@inproceedings{binns_fairness_2018,
	title = {Fairness in {Machine} {Learning}: {Lessons} from {Political} {Philosophy}},
	shorttitle = {Fairness in {Machine} {Learning}},
	url = {https://proceedings.mlr.press/v81/binns18a.html},
	language = {en},
	urldate = {2024-11-29},
	booktitle = {Proceedings of the 1st {Conference} on {Fairness}, {Accountability} and {Transparency}},
	publisher = {PMLR},
	author = {Binns, Reuben},
	month = jan,
	year = {2018},
	note = {ISSN: 2640-3498},
	pages = {149--159},
}

@article{borgatti_network_2011,
	title = {On {Network} {Theory}},
	volume = {22},
	issn = {1526-5455},
	url = {https://doi.org/10.1287/orsc.1100.0641},
	doi = {10.1287/orsc.1100.0641},
	number = {5},
	urldate = {2024-11-29},
	journal = {Organization Science},
	author = {Borgatti, Stephen P. and Halgin, Daniel S.},
	month = sep,
	year = {2011},
	pages = {1168--1181},
}

@book{rogers_diffusion_2003,
	title = {Diffusion of {Innovations}, 5th {Edition}},
	isbn = {978-0-7432-5823-4},
	language = {en},
	publisher = {Simon and Schuster},
	author = {Rogers, Everett M.},
	month = aug,
	year = {2003},
	note = {Google-Books-ID: 9U1K5LjUOwEC},
}

@article{zhao_effects_2011,
	title = {Effects of {Social} and {Temporal} {Distance} on {Consumers}' {Responses} to {Peer} {Recommendations}},
	volume = {48},
	issn = {0022-2437},
	url = {https://doi.org/10.1509/jmkr.48.3.486},
	doi = {10.1509/jmkr.48.3.486},
	language = {en},
	number = {3},
	urldate = {2024-11-30},
	journal = {Journal of Marketing Research},
	author = {Zhao, Min and Xie, Jinhong},
	month = jun,
	year = {2011},
	note = {Publisher: SAGE Publications Inc},
	pages = {486--496},
}

@paper{ICWSM09154,
	author = {Mathieu Bastian and Sebastien Heymann and Mathieu Jacomy},
	title = {Gephi: An Open Source Software for Exploring and Manipulating Networks},
	conference = {International AAAI Conference on Weblogs and Social Media},
	year = {2009},
	url = {http://www.aaai.org/ocs/index.php/ICWSM/09/paper/view/154}
}

@article{fruchterman_graph_1991,
	title = {Graph drawing by force-directed placement},
	volume = {21},
	copyright = {Copyright © 1991 John Wiley \& Sons, Ltd},
	issn = {1097-024X},
	url = {https://onlinelibrary.wiley.com/doi/abs/10.1002/spe.4380211102},
	doi = {10.1002/spe.4380211102},
	language = {en},
	number = {11},
	urldate = {2024-11-30},
	journal = {Software: Practice and Experience},
	author = {Fruchterman, Thomas M. J. and Reingold, Edward M.},
	year = {1991},
	note = {\_eprint: https://onlinelibrary.wiley.com/doi/pdf/10.1002/spe.4380211102},
	pages = {1129--1164},
}

@article{sunstein_law_2002,
	title = {The {Law} of {Group} {Polarization}},
	volume = {10},
	copyright = {2002 Blackwell Publishers Ltd.},
	issn = {1467-9760},
	url = {https://onlinelibrary.wiley.com/doi/abs/10.1111/1467-9760.00148},
	doi = {10.1111/1467-9760.00148},
	language = {en},
	number = {2},
	urldate = {2024-11-29},
	journal = {Journal of Political Philosophy},
	author = {Sunstein, Cass R.},
	year = {2002},
	note = {\_eprint: https://onlinelibrary.wiley.com/doi/pdf/10.1111/1467-9760.00148},
	pages = {175--195},
}

@article{jacomy_forceatlas2_2014,
	title = {{ForceAtlas2}, a {Continuous} {Graph} {Layout} {Algorithm} for {Handy} {Network} {Visualization} {Designed} for the {Gephi} {Software}},
	volume = {9},
	issn = {1932-6203},
	url = {https://journals.plos.org/plosone/article?id=10.1371/journal.pone.0098679},
	doi = {10.1371/journal.pone.0098679},
	language = {en},
	number = {6},
	urldate = {2024-12-01},
	journal = {PLOS ONE},
	author = {Jacomy, Mathieu and Venturini, Tommaso and Heymann, Sebastien and Bastian, Mathieu},
	month = jun,
	year = {2014},
	note = {Publisher: Public Library of Science},
	pages = {e98679},
}

@inproceedings{bakshy_everyones_2011,
	address = {New York, NY, USA},
	series = {{WSDM} '11},
	title = {Everyone's an influencer: quantifying influence on twitter},
	isbn = {978-1-4503-0493-1},
	shorttitle = {Everyone's an influencer},
	url = {https://dl.acm.org/doi/10.1145/1935826.1935845},
	doi = {10.1145/1935826.1935845},
	urldate = {2024-12-09},
	booktitle = {Proceedings of the fourth {ACM} international conference on {Web} search and data mining},
	publisher = {Association for Computing Machinery},
	author = {Bakshy, Eytan and Hofman, Jake M. and Mason, Winter A. and Watts, Duncan J.},
	month = feb,
	year = {2011},
	pages = {65--74},
}

@InProceedings{SciPyProceedings_11,
	author =       {Aric A. Hagberg and Daniel A. Schult and Pieter J. Swart},
	title =        {Exploring Network Structure, Dynamics, and Function using NetworkX},
	booktitle =   {Proceedings of the 7th Python in Science Conference},
	pages =     {11 - 15},
	address = {Pasadena, CA USA},
	year =      {2008},
	editor =    {Ga\"el Varoquaux and Travis Vaught and Jarrod Millman},
}

@article{hu_visualizing_2015,
	title = {Visualizing large graphs},
	volume = {7},
	copyright = {© 2015 Wiley Periodicals, Inc.},
	issn = {1939-0068},
	url = {https://onlinelibrary.wiley.com/doi/abs/10.1002/wics.1343},
	doi = {10.1002/wics.1343},
	language = {en},
	number = {2},
	urldate = {2024-12-11},
	journal = {WIREs Computational Statistics},
	author = {Hu, Yifan and Shi, Lei},
	year = {2015},
	note = {\_eprint: https://onlinelibrary.wiley.com/doi/pdf/10.1002/wics.1343},
	pages = {115--136},
}
\end{sloppypar}
\end{document}